\documentclass[12pt]{article}
\usepackage{amsmath,scalerel}
\usepackage{amsthm}
\usepackage{enumerate}
\usepackage{enumitem}
\usepackage{url}
\usepackage{tikz-cd}
\usepackage{mathtools}

\usepackage{amssymb}
\usepackage{graphics}
\usepackage{float}
\usepackage{epsfig}
\usepackage{epstopdf}
\usepackage{color}

\usepackage{hyperref}
\hypersetup{
    colorlinks=true, %set true if you want colored links
    linktoc=all,     %set to all if you want both sections and subsections linked
    linkcolor=blue,  %choose some color if you want links to stand out
    linktocpage
}

\def\dsp{\def\baselinestretch{1.5}\large\normalsize}
\dsp
\allowdisplaybreaks
\begin{document}
\title{Assessing skin thermal injury risk in exposure tests of heating until flight}
\author{Hongyun Wang\footnote{Corresponding author, hongwang@ucsc.edu} 
\\
Department of Applied Mathematics \\
University of California, Santa Cruz, CA 95064, USA\\
\\
%Wesley A. Burgei,  
Shannon E. Foley \\
U.S. Department of Defense \\
Joint Intermediate Force Capabilities Office \\
Quantico, VA 22134, USA\\
\\
Hong Zhou \\ Department of Applied Mathematics\\
Naval Postgraduate School, Monterey, CA 93943, USA
}

\maketitle
\begin{abstract}
We assess the skin thermal injury risk in the situation where a test subject is exposed to an electromagnetic beam until the occurrence of flight action. The physical process is modeled as follows. The absorbed electromagnetic power increases the skin temperature. Wherever it is above a temperature threshold, thermal nociceptors are activated and transduce an electrical signal. When the activated skin volume reaches a threshold, the flight signal is initiated. After the delay of human reaction time, the flight action is materialized when the subject moves away or the beam power is turned off. The injury risk is quantified by the thermal damage parameter calculated in the Arrhenius equation. It depends on the beam power density absorbed into the skin, which is not measurable. In addition, the volume threshold for flight initiation is unknown. To circumference these difficulties, we normalize the formulation and write the thermal damage parameter in terms of the occurrence time of flight action, which is reliably observed in exposure tests. This thermal injury formulation provides a viable framework for investigating the effects of model parameters. 
\end{abstract}


%%%
%%%
%%%
\section{Introduction}
Millimeter-wave (MMW) systems in the range of 30-300 GHz are becoming increasingly important in both defense and civilian applications, such as wireless communications, security scanning, modern medicine, industrial heating devices, police traffic devices, and military devices. These wide applications prompt a better understanding of the biological effects of MMW exposures.  

One major concern of such systems is the safety of human exposure to MMWs. Earlier clinical and experimental studies into the biological effects of MMW exposure found that the principal impact of MMWs is thermal and skin is the primary organ absorbing the MMW energy 
\cite{Romanenko_2017, Zhadobov_2011, Quement_2014}. The MMW energy penetrates less than 1mm into the skin and causes the skin temperature to increase by dielectric heating.   

The local temperature rise of skin depends on several factors, including 
the intensity and spot size of the MMW radiation, duration of the exposure, 
the absorption coefficient for the MMW frequency, heat conductivity of skin, 
heat exchange via blood perfusion, and heat loss at skin surface (via black body radiation, air convection, sweating) \cite{Nelson_2000}. 
Skin thermal damage occurs not only due to a high temperature but also due to the time duration of high temperature.. There are two potential thermal hazards from millimeter waves: thermal pain and superficial burn. For example, in the 94 GHz exposure with a 1.8 $\text{W/cm}^2$ beam, the thermal pain starts when the surface temperature 
reaches approximately $44^\circ$C, about $10^\circ$C increase over a typical skin baseline temperature \cite{Walters_2000}. In contrast, the threshold for burn corresponded to peak temperatures of approximately $60^\circ$C, and required a beam energy fluence
(energy delivered per area) 2-3 times higher than that producing 
thermal pain \cite{Foster_2010}.  

Most MMW systems use low powers. However, certain commercial and military applications have MMW systems operate at high powers. In these scenarios there is a possibility of accidental over-exposure to MMWs that may cause thermal injury. In the present study we focus on the effect of a high power MMW on skin tissues in the situation where the exposure lasts until the occurrence of flight. 
We investigate how the human reaction time, the MMW beam specifications, and other parameters affect skin thermal damage. 

This paper is organized into four subsequent sections. Section 2 describes the mathematical 
formulation for modeling skin thermal damage caused by a millimeter-wave radiation. 
In Section 3, we list numerical values collected from published literature for all parameters, 
except the unknown volume threshold for flight initiation. 
We establish a normalized formulation to work around the unknown volume threshold. 
This is followed by numerical simulations in Section 4, investigating the thermal damage
vs the observed time of flight action and its sensitivity with respect to model parameters. 
Finally, in Section 5, we discuss the results and conclusions obtained. 

%%%
%%%
%%%
\section{Mathematical formulation} \label{formulation}
\subsection{Overview of the model} 
We consider the situation where a skin area of a test subject is exposed to 
a millimeter wave radiation. The absorbed electromagnetic power becomes a heat source, 
increasing the skin temperature. Wherever the local temperature is above the activation 
threshold, thermal nociceptors are activated and transduce an electrical signal, 
which is transmitted to the brain by the nervous system. 
When the burn sensation (the aggregated electrical signal from all thermal nociceptors 
received in the brain) exceeds a subjective threshold, the brain issues a flight instruction signal, 
which propagates via the nervous system to the muscles. Finally, the muscles carry out 
the flight in the form of the subject moving away from the electromagnetic beam and/or 
turning off the beam power. 

The flight is initiated when a sufficiently large electrical signal is transduced 
at the exposed skin spot. 
The flight action is eventually materialized after the time delay of the nociceptor signal 
traveling from the exposed spot to the brain, the transduction of 
flight instruction signal in the brain and the flight signal traveling to the muscles.
This time delay between the flight initiation and the flight actuation is referred to as 
the simple reaction time (SRT) or the human reaction time. 
The beam radiation on the skin is terminated only upon the actuation of flight 
when the test subject moves away from the beam or turns off the beam power. 
The beam radiation lasts until the actuation of flight. It does not end at the flight initiation. 
Because of the delay in power off, the human reaction time affects the thermal injury risk.

The flight initiation is determined by the aggregated nociceptor signal transduced 
at the exposed skin spot and a subjective threshold. In this study, we consider 
the idealized situation where thermal nociceptors are uniformly distributed in skin tissue. 
As a result, the aggregated nociceptor signal is proportional 
to the activated skin volume where the local temperature is above the nociceptor 
activation threshold. 
Under this assumption, the flight initiation is determined by the activated skin volume
and a corresponding volume threshold. 

For a high power MMW, while the beam power is on, the skin temperature increases 
monotonically driven by the absorbed electromagnetic power where the cooling effect of 
blood perfusion and surface heat loss are relatively negligible.
As a result, the activated skin volume increases monotonically and the flight initiation 
threshold will be reached as the exposure continues. 
We study the skin thermal damage risk in the experimental setting where 
the beam radiation lasts until the occurrence of flight action. 
Upon the flight action, the active beam heating if off and the spatial maximum skin 
temperature starts decreasing.  
However, the skin temperature remains high for a period of time after the end of beam power. 
To properly assess the skin thermal damage risk, we need to follow the temperature 
evolution well beyond the end of beam power, until the temperature drops back close 
to the skin baseline temperature. 
From the skin temperature, we use the Arrhenius equation to calculate 
the thermal damage parameter $\Omega $. 
To make the study relevant for real exposure tests, 
we need a practically measurable quantity to parameterize the thermal damage risk. 
In exposure tests, the beam power density absorbed into the skin is 
a significant contributing factor of thermal injury. This quantity, however, is not 
measurable in exposure tests. In contrast, the occurrence time of flight action 
is reliably observed in exposure tests. Thus, we use the time of flight action 
to parameterize the thermal damage risk. 
\subsection{The coordinate system}
We first establish the coordinate system as in our previous study \cite{Wang_2020}. 
We consider the case of a flat skin surface. 
The normal direction of the skin surface going into the skin tissue is selected as 
the positive $z$-direction. The skin surface is $z=0$ and $z > 0$ represents 
the skin tissue. 
In a plane perpendicular to the $z$-axis (i.e., with $z = z_0$), 
the 2D coordinates are denoted by $\mathbf{r}=(x, y)$. 
We consider the situation where the incident electromagnetic beam is 
perpendicular to the skin surface (i.e., along the $z$-direction). 
The center of the incident beam is the location of spatial maximum power density
over the beam cross section. 
On the skin surface, we set the beam center as the origin 
$\mathbf{r} = (0, 0)$ of the $xy$-plane. 
In the skin tissue, the 3D coordinates of a point are represented by 
$(\mathbf{r}, z) = (x, y, z), \; z \ge 0$. 

\subsection{Physical quantities and equations}
We introduce the physical quantities and equations used in our model. 
\begin{itemize}
\item $(\mathbf{r}, z) = (x, y, z), \; z \ge 0$ denotes the 3D coordinates of a point in the skin tissue.
\item $T(\mathbf{r}, z, t)$ is the skin temperature at position $(\mathbf{r}, z)$ at time $t$.
\item $\rho_m$ is the mass density, $C_p$ the specific heat capacity, and 
$k$ the heat conductivity of skin. 
In this study, we consider a single layer of uniform skin where 
all skin material properties are independent of position $(\mathbf{r}, z)$.
\item $P_d(\mathbf{r})$ is the beam power density passing through the skin 
surface at position $\mathbf{r}=(x, y)$. 
In our model, $P_d(\mathbf{r})$ is the power density that is absorbed 
into the skin tissue (and becomes heat). 
It has the expression $P_d(\mathbf{r}) = \alpha P_d^\text{(i)}(\mathbf{r})$ 
where $P_d^\text{(i)}(\mathbf{r})$ is the incident beam power density and 
$\alpha $ is the fraction of beam power passing through the skin surface
(the rest is reflected). 
In exposure tests, $P_d(\mathbf{r}) $ is not measured. For that reason, 
we shall not treat $P_d(\mathbf{r}) $ as a measurable quantity. 
\item $\mu$ is the skin's absorption coefficient for the
electromagnetic frequency as it propagates in skin.
$\mu $ is different from coefficient $\alpha $ introduced above. 
$\alpha $ is the fraction of beam power passing through the skin surface.
$\mu $ is the fraction of beam power absorbed \underline{per depth}
as the beam propagates in the depth of skin. 
$\mu$ has the physical dimension of $1/[\text{length}]$ 
while $\alpha $ is dimensionless.  
\item 
$p_v(\mathbf{r}, z)$ is the power absorbed \underline{per volume} at position
$(\mathbf{r}, z)$, which provides the heating power
per volume in the temperature evolution. 
$p_v(\mathbf{r}, z)$ decays exponentially in the depth direction 
and has the expression 
\begin{equation}
p_v(\mathbf{r}, z) 
= P_d(\mathbf{r}) \mu e^{-\mu z}
\label{pv_exp}
\end{equation}
\item
\underline{The balance of energy} gives the governing equation for $T(x, y, z, t)$.
\begin{equation}
\underbrace{\rho_m C_p\frac{\partial T}{\partial t}}_{\substack{
\text{rate of change}\\ \text{of heat in skin}}}
= \underbrace{k \Big(\frac{\partial^2T}{\partial x^2}
+\frac{\partial^2T}{\partial y^2}+\frac{\partial^2T}{\partial z^2}
\Big)}_{\substack{
\text{net heat influx}\\ \text{from conduction}}}
+ \underbrace{P_d(\mathbf{r})\mu e^{-\mu z}}_\text{absorbed power}
\label{heat_Eq1}
\end{equation}
\item
\underline{Effect of lateral heat conduction}. 
In most exposure tests, the length scale in the lateral dimensions is much larger than 
the penetration depth of the electromagnetic frequency into the skin tissue. 
In this situation, to the leading term approximation, the lateral heat conduction 
is negligible. The governing equation becomes 
\begin{equation}
\rho_m C_p\frac{\partial T}{\partial t}= k \frac{\partial^2T}{\partial z^2}
+ P_d(\mathbf{r})\mu e^{-\mu z}
\label{heat_Eq2}
\end{equation}
\item 
$T_\text{base}$ is the baseline temperature. 
We assume that the skin temperature is uniform in $(x, y, z)$ at the start of beam exposure
($t = 0$), which is called the baseline temperature and is denoted by $T_\text{base}$. 
It gives the initial condition for temperature evolution: $T(x, y, z, 0) = T_\text{base}$.
\item 
\underline{Zero heat flux at the skin surface}. 
At the skin surface, body heat is slowly lost to the environment 
via the black body radiation and the convection cooling by the surrounding air. 
In homeostasis, this heat loss is countered by the heat influx carried by blood perfusion. 
For a high power MMW exposure, the heat loss at the surface is negligible 
relative to the large energy influx (the beam power passing into skin). 
In our study, we neglect the slow heat loss at the skin surface. 
For the same reason, the effect of blood perfusion is also neglected in the 
energy balance of (\ref{heat_Eq2}). 
Zero heat flux at skin surface leads to the boundary condition 
$\partial T/\partial z \big|_{z=0} = 0$. 
\item 
\underline{Semi-infinite skin tissue.}
Given the exponential decay of $p_v(\mathbf{r}, z)$ in (\ref{pv_exp}), 
the beam power is virtually all absorbed within a few multiple of 
$1/\mu$, which is the penetration depth of electromagnetic wave into the skin tissue. 
In our study, $1/\mu = 0.16\, \text{mm}$. 
For the purpose of modeling the temperature evolution, we can approximately treat 
the skin tissue as a semi-infinite body, mathematically extending to $z=+\infty$. 
The temperature is governed by the initial boundary value problem (IBVP) 
\begin{equation} 
\begin{dcases}
\rho_m C_p\frac{\partial T}{\partial t}=k \frac{\partial^2T}{\partial z^2} 
+ P_d(\mathbf{r})\mu e^{-\mu z} \\[1ex]
\frac{\partial T(x, y, z, t)}{\partial z} \bigg|_{z=0}=0 \\
T(x, y, z, 0) = T_\text{base}
\end{dcases} 
\label{IBVP_1}
\end{equation}
\item 
$T_\text{act}$ is the activation temperature of thermal nociceptors. 
\item
$v_\text{act}(t)$ is the activated skin volume at time $t$. 
It has the expression 
\begin{equation}
v_\text{act}(t) \equiv \text{volume}\Big\{(x, y, z) 
\Big| T(x, y, z, t) \ge T_\text{act} \Big\}
\label{v_act}
\end{equation}
Here $v_\text{act}(t)$ has dependence on $P_d(\mathbf{r}) $ via $T(x, y, z, t)$.
\item 
$v_c$ is the threshold on the activated skin volume for flight initiation. 
In this study, $v_c$ is unknown but is deterministic 
(i.e., not fluctuating from one test to another). 
\item 
$t_c $ is the time of flight initiation when the activated 
skin volume reaches the threshold $v_c$. Given $v_c$, 
$t_c$ is solved from the equation 
\begin{equation}
v_\text{act}(t) \Big|_{t=t_c} = v_c
\label{t_c}
\end{equation}
Here $t_c$ has dependence on $P_d(\mathbf{r}) $ via 
$ v_\text{act}(t) $. 
\item
$t_R$ is the human reaction time, also called the simple reaction time (SRT). 
\item
$t_F $ is the time of flight actuation, which is the sum of $t_c$ and $t_R$.
\begin{equation}
t_F = t_c + t_R
\label{t_F}
\end{equation}
Here $t_F$ has dependence on $P_d(\mathbf{r})$ via $t_c$. 
Note that while the flight is initiated at time $t_c$, the actuation of flight is delayed 
by the human reaction time $t_R$. The electromagnetic heating continues all the way until 
$t_F$ when the test subject moves away from the beam or turns off the beam power. 
\item 
\underline{The Arrhenius equation} describes 
the reaction rate constant of thermal damage 
at location $(x, y, z)$ at time $t$. It is based on the local temperature 
$T(x, y, z, t)$. 
$$ k_\text{d}(x, y, z, t) = A \exp\big(\frac{-\Delta E_a}{R \, T(x, y, z, t)} \big) $$
The integral of the thermal damage rate constant over time $t$ gives 
the dimensionless thermal damage parameter, $\Omega$, which indicates the degree 
of burn injury. 
Since the region near beam center on the skin surface has the highest temperature, 
we focus on that region when assessing the skin thermal damage. 
\begin{equation}
\Omega = \int_0^{t_\text{stl}} A \exp\big(
\frac{-\Delta E_a}{R \, T(t)} \big) dt 
\label{Arrhenius_Eq}
\end{equation}
Here $T(t)$ is the skin surface temperature at the beam center at time $t$ 
in Kelvin (K). In the integral, the lower limit is the starting time of beam power; 
the upper limit, $t_\text{stl}$, is the settlement time when  
the skin temperature settles back close to the baseline temperature after 
the beam power is off. 
The thermal damage parameter $\Omega$ has dependence on 
$P_d(\mathbf{r})$ via $T(t)$.
\item 
\underline{Parameters in the Arrhenius equation.}
Parameter $A$ is the frequency factor representing 
molecular collisions for exciting protein denaturation;  
$\Delta E_a $ is the activation energy barrier for protein denaturation;
and $R$ is the ideal gas constant. 
\end{itemize}
%

%%%
%%%
%%%
\section{Parameter values and normalized formulation}
\subsection{Parameter values and derived quantities} \label{para_values}
Below we list the parameter values collected from published literature with sources. 
Some of parameter values are the same as those used in 
the ADT CHEETEH model \cite{Cazares_2019}. 
\begin{itemize}
\item Mass density of skin (dermis): $\rho_m =  1116\, \text{kg/m}^3$ 
\cite{Duck_1990, Xu_2010}. 
\item Specific heat capacity of skin (dermis): $C_p =  3300\, \text{J/(kg K)}$ 
\cite{Xu_2010, Henriques_1947}. 
\item Heat conductivity of skin (dermis): $k = 0.445\, \text{W/(m K)}$ 
\cite{Xu_2010, Elkins_1973}. 
\item 
For a Gaussian beam, the power density has the form 
\begin{equation}
P_d(\textbf{r}) = P_d^{(0)} \exp\Big( \frac{-2 |\textbf{r}|^2}{r_b^2} \Big) 
\label{Gauss_beam}
\end{equation}
where $r_b$ is the beam radius and $P_d^{(0)}$ is the beam center power density 
absorbed into skin. $P_d^{(0)}$ is not measurable in exposure tests. 
\item Penetration depth of a 94 GHz beam into skin: $1/\mu = 0.16\, \text{mm}$ 
\cite{Walters_2020}. 
\item 
Baseline temperature: $T_\text{base} = 32\, ^\circ\text{C} = 305.15\, \text{K}$ 
\cite{ Cazares_2019, Olesen_1982}. 
\item
$T(x, y, z, t)$, the 3D temperature distribution of skin at time $t$, 
is \underline{a derived quantity}. 
It is solved from IBVP (\ref{IBVP_1}). $T(x, y, z, t)$ depends on  
$P_d(\mathbf{r}) $.
\item 
Activation temperature of thermal nociceptors: 
$T_\text{act} = 40.4\, ^\circ\text{C} = 313.55\, \text{K}$
\cite{Tillman_1995}.
\item
$v_\text{act}(t)$, the activated skin volume at time $t$, 
is \underline{a derived quantity}. 
It is calculated from $T(x, y, z, t)$ using formula (\ref{v_act}). 
$v_\text{act}(t)$ has dependence on $P_d(\mathbf{r}) $ via $T(x, y, z, t)$.
\item 
$v_c$, the threshold on the activated skin volume for flight initiation, 
is \underline{unknown}. 
\item 
$t_c$, the time of flight initiation, is  \underline{a derived quantity}. It is solved 
in equation $v_\text{act}(t_c) = v_c$. 
$t_c$ has dependence on $P_d(\mathbf{r})$ via $v_\text{act}(t)$. 
\item
Simple reaction time (SRT): $t_R = 275\, \text{ms}$
\cite{Dykiert_2012, Woods_2015}. The mean SRT of females (281 ms)
is slightly longer than that of males (268 ms).
\item
$t_F = t_c + t_R$, the time of flight action, 
is \underline{a derived quantity}. 
$t_F $ is the time when the beam power is turned off and it 
depends on $P_d(\mathbf{r})$ via $t_c$. 
\item 
\underline{Parameters in the Arrhenius equation:} 
the frequency factor is $A = 8.82\times 10^{94}\, \text{s}^{-1}$; 
the activation energy barrier for protein denature is 
$\Delta E_a = 6.03\times 10^5\, \text{J/mol} $; 
the ideal gas constant is $R = 8.314\, \text{J/(mol K)} $
\cite{Pearce_2010, Pearce_2017}. 
\item 
$\Omega $, the dimensionless thermal damage parameter, is \underline{a derived quantity}. 
$\Omega $ is calculated in formula (\ref{Arrhenius_Eq}) from the skin temperature 
$\{T(t)\}$ over the time course from the start of beam power to the settlement time 
$t_\text{stl}$ when the skin temperature \underline{settles} back close to the baseline temperature after the beam power is off.  
\item 
\underline{Classifications of burn injury}. Burn injuries are classified into three main categories: first, second, and third-degree burns, depending on
how deep and how severely they affect the skin tissue. 
First-degree (or superficial)  burns only affect the outer layers of the skin without
long-term damage. Second-degree (or partial-thickness) burns penetrate deeper than first-degree burns and usually take longer to heal. 
Third-degree (or full-thickness) burns can cause very serious injuries and require extensive rehabilitation. 
We adopt the standard classification below, based on the magnitude of 
thermal damage parameter $\Omega $ \cite{Pearce_2017, Orgill_1998}.
\begin{equation}
\begin{dcases} 
0.53 \le \Omega < 1 &  \longrightarrow \;\; \text{first-degree burn} \\
1 \le \Omega < 10^4 & \longrightarrow \;\; \text{second-degree burn} \\
\Omega \ge 10^4 &  \longrightarrow \;\; \text{third-degree burn} 
\end{dcases} 
\label{burn_class} 
\end{equation}
\end{itemize}

\subsection{Normalized formulation} \label{norm_form}
Among all parameters listed in subsection \ref{para_values}, two are unknown
or not measurable in exposure tests:
1) the volume threshold for flight initiation ($v_c$) and 
2) the beam center power density absorbed into skin ($P_d^{(0)}$). 
The effect of $P_d^{(0)}$ is well captured in the occurrence time of flight action $t_F$,
which is observed in exposure tests. 
To get around the unknown volume threshold $v_c$, we select scales 
for normalization as follows.

The penetration depth of 94 GHz wave into skin provides a 
length scale in the depth direction, which we denote by $z_s$.
\begin{equation}
z_s \equiv 1/\mu = 0.16\, \text{mm} 
\nonumber 
\end{equation}
To get rid of the unknown volume threshold $v_c$ in the normalized formulation, 
we select the lateral length scale $r_s$ as  
\begin{equation}
r_s \equiv \sqrt{\frac{v_c}{\pi z_z} } = \sqrt{\frac{\mu v_c}{\pi } }
= \text{unknown} 
\nonumber 
\end{equation}
Geometrically, $r_s$ is the radius of the cylinder with volume = $v_c$ and 
height = $z_s$. In this way, after normalization, the non-dimensional volume threshold 
is $\pi$. 

The time scale is given by the characteristic time of heat diffusion by a distance
of $z_s$ in the depth direction. The lateral heat diffusion is negligible. 
\begin{equation}
t_s \equiv \frac{\rho_\text{m} C_p}{k} z_s^2
= \frac{\rho_\text{m} C_p}{k \mu^2} = 0.2119 s
\label{t_s}
\end{equation}
In the model formulation, we normalize $(\mathbf{r}, z, t) $ 
using scales $(r_s, z_s, t_s) $.
\begin{align}
& z_s \equiv 1/\mu = 0.16\, \text{mm}, \qquad z_\text{n} = \frac{z}{z_s} = \mu z
\label{z_nd} \\
& t_s \equiv \frac{\rho_\text{m} C_p}{k \mu^2} = 0.2119 s, 
\qquad t_\text{n} = \frac{t}{t_s}
\label{t_nd} \\
& r_s \equiv \sqrt{\frac{\mu v_c}{\pi } }, \qquad 
\mathbf{r}_\text{n} = \frac{\mathbf{r}}{r_s}, \qquad
r_\text{b,n} = \frac{r_b}{r_s} , \qquad
v_\text{c,n} = \frac{v_c}{z_s r_s^2} = \pi  
\label{r_nd}
\end{align} 
The temperature is unchanged. As a function of 
$(x_\text{n}, y_\text{n}, z_\text{n}, t_\text{n})$. it is denoted by $T^\text{(n)}$.
$$ T^\text{(n)}(x_\text{n}, y_\text{n}, z_\text{n}, t_\text{n} ) 
= T(x, y, z, t) $$
After normalization, the governing equation of $T^\text{(n)}$ has the same form as IBVP (\ref{IBVP_1}), 
with coefficients updated.
\begin{equation} 
\begin{dcases}
 \frac{\partial T^\text{(n)}}{\partial t_\text{n}} = 
 \frac{\partial^2T^\text{(n)}}{\partial z_\text{n}^2 } 
+ P_\text{d}^\text{(n)}(\mathbf{r}_\text{n}) e^{-z_\text{n}} \\[1ex]
\frac{\partial T^\text{(n)}(x_\text{n}, y_\text{n}, z_\text{n}, t_\text{n})}{\partial z_\text{n} } 
\bigg|_{z_\text{n}=0}=0 \\
T^\text{(n)}(x_\text{n}, y_\text{n}, z_\text{n}, 0) = T_\text{base}
\end{dcases} 
\label{IBVP_1B}
\end{equation}
where the normalized beam power density has the expression 
\begin{equation}
P_\text{d}^\text{(n)}(\mathbf{r}_\text{n}) = \frac{1}{k \mu} P_d(\mathbf{r}) 
= \frac{P_d^{(0)}}{k \mu} \exp\Big( \frac{-2 |\textbf{r}_\text{n} |^2}{r_\text{b,n}^2} \Big) 
\label{P_d_n}
\end{equation}
Note that both $T^\text{(n)}$ and $P_\text{d}^\text{(n)}$ 
have the physical dimension of temperature while 
the normalized beam radius $r_\text{b,n}$ is dimensionless. 
Given parameters $(T_\text{base}, P_d^{(0)}, k, \mu, r_\text{b,n}) $, 
we solve (\ref{IBVP_1B}) with (\ref{P_d_n}) to calculate 
the temperature distribution 
$T^\text{(n)}(x_\text{n}, y_\text{n}, z_\text{n}, t_\text{n} )$. 
Then we find the activated volume in the normalized variables 
$(x_\text{n}, y_\text{n}, z_\text{n}, t_\text{n})$. 
\begin{equation}
v_\text{act,n}(t_n) \equiv \text{volume}\Big\{(x_n, y_n, z_n) 
\Big| T^\text{(n)}(x_n, y_n, z_n, t_n) \ge T_\text{act} \Big\}
\label{v_act_n}
\end{equation}
The activated volume in $(x, y, z, t)$ and that in 
$(x_\text{n}, y_\text{n}, z_\text{n}, t_\text{n})$
are related by 
\begin{equation}
v_\text{act}(t) = (r_s^2 z_s) v_\text{act,n}(t_n) = \frac{v_c}{\pi} v_\text{act,n}(t_n)
\label{v_act_2}
\end{equation}
Recall that the flight initiation is governed by 
$v_\text{act}(t) \big|_{t=t_c} = v_c$, which becomes 
\begin{equation}
v_\text{act,n}(t_n) \Big|_{t_n = t_\text{c,n}}= \pi 
\label{v_threshold} 
\end{equation}
Given the normalized activated volume $v_\text{act,n}(t_n)$, 
equation (\ref{v_threshold}) is independent of the unknown volume threshold $v_c$. 
The effect of unknown $v_c$ is hidden in $v_\text{act,n}(t_n)$ 
via the normalized beam radius $r_\text{b,n} \equiv \frac{r_b}{r_s}$
where the lateral length scale $r_s$ contains $v_c$. 
We specify the beam radius as a multiple of $r_s$ (i.e., specifying 
the dimensionless $r_\text{b,n}$), instead of giving a value with a physical unit
With this setup, the entire normalized formulation is independent of 
the unknown $v_c$. 

In exposure tests, the beam power is turned off at the time of flight action, 
delayed by the human reaction time $t_R$ from the flight initiation. 
To assess the thermal damage parameter in (\ref{Arrhenius_Eq}), 
we need to follow the skin temperature from the start of beam power 
to the settlement time $t_{stl}$ when the skin temperature drops 
back to close to the baseline temperature. 
In the normalized formulation, we work with the normalized time 
$t_\text{n}$ and $T^\text{(n)}$ vs $t_n$. The normalized time is related to 
the physical time by the time scale $t_s$: $ t_\text{n} = \frac{t}{t_s}$. 
We use change of variables to write $\Omega$ 
in terms of function $T^\text{(n)}(t_n)$. 
\begin{align}
\Omega & = t_s \int_0^{t_\text{stl,n}} A \exp\big(
\frac{-\Delta E_a}{R \, T^\text{(n)}(t_\text{n})} \big) dt_\text{n} 
\label{Arrhenius_Eq_n}
\end{align}
The degree of burn injury is determined from $\Omega $ according to (\ref{burn_class}). 

In summary, we represent the beam radius as a multiple of the unknown length scale 
$r_s$, which is derived from the unknown volume threshold $v_c$. 
The resulting normalized formulation is independent of the unknown $v_c$. 
This allows us to carry out simulations to assess burn injury risk without 
knowing the value of the volume threshold $v_c$. 

\subsection{Practical parameterization of thermal damage risk}
In the previous subsection, we discussed the advantage of working 
in the normalized formulation: all parameters in (\ref{IBVP_1B}) are known 
and are listed in subsection \ref{para_values}, 
except the specifications of the Gaussian beam: $(r_\text{b}, P_d^{(0)})$. 

To get around the unknown volume threshold $v_c$, in our study the physical 
beam radius $r_b$ is specified as a known multiple of the unknown $r_s$, which 
is derived from the unknown $v_c$. Simulations are carried out 
in the the normalized formulation with the known value of 
$r_\text{b,n} = \frac{r_b}{r_s}$. 
Scaling with the unknown $r_s$ allows computing the thermal damage parameter 
$\Omega $ and assessing the degree of burn injury. 

Hyperthetically, we could specify $P_d^{(0)}$ in theoretical simulations 
and use it to parameterize the corresponding thermal damage risk.
$P_d^{(0)}$ is the beam center power density \underline{absorbed into the skin tissue}, 
which is affected by the propagation loss in air and the reflectance of the skin surface. 
Operationally, $P_d^{(0)}$ is not measurable in exposure tests. 
Using $P_d^{(0)}$ to parameterize the thermal damage risk 
is not practically meaningful in real exposure tests. 
In this study, we use the occurrence time of flight action $t_F$  to parameterize the 
thermal damage risk. In real exposure tests, $t_F$ is directly observed 
as the time when the test subject moves away from the beam radiation 
or turns off the beam power. 
In  (\ref{v_act}), the activated volume 
$v_\text{act}(t)$ increases with time $t$ at any fixed beam power density $P_d^{(0)}$ 
and increases with $P_d^{(0)}$ at any fixed $t$. 
As a result, in (\ref{t_c}) a higher value of $P_d^{(0)}$ implies that
the volume threshold $v_c$ is attained at a smaller time of flight initiation $t_c$. 
That is, as a function, $t_c = g(P_d^{(0)})$ is monotonically decreasing
and thus is invertible. 
This monotonic trend allows using $t_F$ instead of $P_d^{(0)}$ to parameterize 
the thermal damage risk. For a given observed $t_F$, the underlying $P_d^{(0)}$ 
is determined as
\begin{equation}
P_d^{(0)} = g^{-1}(t_F-t_R) 
\label{P_d_sol}
\end{equation}

In our study, we calculate the thermal damage parameter $\Omega $ 
as a function of $(t_F, r_b/r_s)$. In this setting, the absorbed beam center power density 
$P_d^{(0)}$ is hidden in $t_F$; the effect of the unknown volume threshold $v_c$
is hidden in the unknown lateral scale $r_s$. The price of getting rid of the unknown 
$v_c$ is that we need to specify the beam radius as a multiple of the unknown 
length scale $r_s$. 

\subsection{Analytical solution of $T^\text{(n)}(x_\text{n}, y_\text{n}, z_\text{n}, t_\text{n})$} 
We start with the case where the beam power is turned on at $t_\text{n} = 0$ 
and is kept on indefinitely. The temperature as a function of 
normalized variables is given by  
\begin{equation} 
T^\text{(n)}(x_\text{n}, y_\text{n}, z_\text{n}, t_\text{n}) 
= T_\text{base} + 
\frac{P_d^{(0)}}{k \mu} \exp\Big( \frac{-2 |\textbf{r}_\text{n} |^2}{r_\text{b,n}^2} \Big)
U(z_\text{n}, t_\text{n})
\label{Tn_exp}
\end{equation}
where $U(z, t) $ satisfies a parameter-free IBVP. 
\begin{equation} 
\begin{dcases}
 \frac{\partial U}{\partial t} = 
 \frac{\partial^2 U}{\partial z^2 } + e^{-z} \\[1ex]
\frac{\partial U(z, t)}{\partial z} 
\bigg|_{z=0}=0 \\
U(z, 0) = 0
\end{dcases} 
\label{IBVP_1C}
\end{equation}
IBVP (\ref{IBVP_1C}) has a closed-form analytical solution. 
\begin{equation}
U(z, t) = \begin{dcases} 
-e^{-z}+\frac{e^{-z+t}}{2} \text{erfc}(\frac{-z+2t}{\sqrt{4t} }) + \frac{e^{z+t}}{2} \text{erfc}(\frac{z+2t}{\sqrt{4t} }) & \\
\hspace{1in} -z \, \text{erfc}(\frac{z}{\sqrt{4 t} }) 
+ \frac{2\sqrt{t}}{\sqrt{\pi }} e^{\frac{-z^2}{4t}}, & t > 0 \\
0, & t \le 0 
\end{dcases} 
\label{U0_sol} 
\end{equation}
When the beam power is turned off at $t_\text{n} = t_\text{F,n}$
(the time of flight action), 
the temperature is given as the difference between two functions.
\begin{equation} 
T^\text{(n)}(x_\text{n}, y_\text{n}, z_\text{n}, t_\text{n}) 
= T_\text{base} + 
\frac{P_d^{(0)}}{k \mu} \exp\Big( \frac{-2 |\textbf{r}_\text{n} |^2}{r_\text{b,n}^2} \Big)
\Big( U(z_\text{n}, t_\text{n}) - U(z_\text{n}, t_\text{n}-t_\text{F,n} )  \Big) 
\label{Tn_exp2}
\end{equation}

%%%
%%%
%%%
\section{Simulations and results}
\subsection{Fully non-dimensional formulation for simulations}
In our simulations, we work with the fully non-dimensional formulation.
The non-dimensional temperature and non-dimensional beam center power density are 
\begin{align}
& T_\text{nd} \equiv \frac{T^\text{(n)} -T_\text{base} }{T_\text{act} -T_\text{base}}, \qquad 
T_\text{base,nd} = 0, \qquad T_\text{act,nd} = 1 
\label{T_nd} \\[1ex] 
& P_\text{d,nd}^{(0)} \equiv \frac{P_d^{(0)}}{P_\text{s}} , \qquad
P_\text{s} \equiv k \mu (T_\text{act} -T_\text{base}) 
\label{P_nd} 
\end{align}
$z_\text{n} $, $t_\text{n}$ and $\mathbf{r}_\text{n} $  are already non-dimensional
as defined in  (\ref{z_nd}),  (\ref{t_nd}) and (\ref{r_nd}). 
While the beam power is on, the non-dimensional temperature has the expression: 
\begin{align}
& T_\text{nd}(x_\text{n}, y_\text{n}, z_\text{n}, t_\text{n}) = P_\text{d,nd}^{(0)} 
\exp\Big( \frac{-2 |\textbf{r}_\text{n} |^2}{r_\text{b,n}^2} \Big) U(z_\text{n}, t_\text{n})
\label{T_nd_exp}
\end{align}
Here $P_\text{d,nd}^{(0)}$ and $T_\text{nd}$ are non-dimensional and are given
in (\ref{P_nd}) and (\ref{T_nd_exp}). 
In terms of $T_\text{nd}(x_\text{n}, y_\text{n}, z_\text{n}, t_\text{n})$, 
the non-dimensional activated volume, the non-dimensional time of flight initiation, 
and the non-dimensional time of flight action are calculated as 
\begin{align}
& v_\text{act,n}(t_\text{n}) = \text{vol}\Big\{(x_\text{n}, y_\text{n}, z_\text{n}) \Big| T_\text{nd}(x_\text{n}, y_\text{n}, z_\text{n}, t_\text{n}) \ge T_\text{act,nd} \Big\}, 
\qquad T_\text{act,nd} = 1
\nonumber \\
& v_\text{act,n}(t_\text{n}) \Big|_{t_\text{n}=t_\text{c,n}} = v_\text{c,n}, 
\qquad v_\text{c,n} = \pi
\nonumber \\
& t_\text{F,n} = t_\text{c,n} + t_\text{R,n}, 
\qquad t_\text{R,n} = \frac{t_R}{t_s} =\frac{0.275 \text{ s}}{0.2119 \text{ s}}
= 1.298
\nonumber
\end{align}

In exposure tests, the beam power is turned off at the time of flight action $t_\text{F,n}$, 
not at the time of flight initiation. To correctly assess the thermal damage parameter
$\Omega $, we follow the temperature evolution, far beyond the beam power end time, 
until the temperature falls back close to the baseline temperature. 
Over the entire time course, the non-dimensional skin surface temperature 
at the beam center, $T_\text{nd}(t_\text{n}) \equiv T_\text{nd}(0, 0, 0, t_\text{n})$, 
has a unified expression 
\begin{align}
& T_\text{nd}(t_\text{n}) = P_\text{d,nd}^{(0)}  \Big( 
h(t_\text{n}) - h(t_\text{n}-t_\text{F,n} ) \Big) , \quad 
\text{for any $t_\text{n}$} 
\nonumber\\
& h(s) \equiv U(0, s) = \begin{dcases} 
2 \frac{\sqrt{s }}{\sqrt{\pi}} -1 + \text{erfcx}(\sqrt{s }), & s > 0 \\
0, & s \le 0 
\end{dcases} 
\nonumber
\end{align}
From (\ref{Arrhenius_Eq_n}), we write the thermal damage parameter $\Omega $ 
in terms of non-dimensional $T_\text{nd}(t_\text{n})$. 
\begin{align}
& \Omega = t_\text{n} \int_0^{t_\text{stl,n}} A \exp\big(
\frac{-\Delta E_a}{R \, \big[ T_\text{base} + 
(T_\text{act} -T_\text{base}) T_\text{nd}(t_\text{n}) \big] } \big) dt_\text{n} 
\label{Arrhenius_Eq_n2}
\end{align}
For numerical integration, we view it as an integral of a function of $t_\text{n}$ 
\begin{align}
& \Omega = \int_0^{t_\text{stl,n}} k_\text{d}(t_\text{n}) dt_\text{n}, 
\qquad k_\text{d}(t_\text{n}) \equiv c_1 \exp\big(
\frac{-1}{c_2 + c_3  T_\text{nd}(t_\text{n}) } \big)
\label{Omega_calc} \\
& \quad c_1 \equiv t_s A = \frac{\rho_\text{m} C_p}{k \mu^2} A = 1.869\times 10^{94}
\nonumber \\
& \quad c_2 \equiv \frac{R}{\Delta E_a} T_\text{base} = 4.207\times 10^{-3} , \quad 
c_3 \equiv \frac{R}{\Delta E_a} (T_\text{act}-T_\text{base}) 
= 1.158\times 10^{-4}
\nonumber 
\end{align}
All quantities in (\ref{Omega_calc}) are non-dimensional, including coefficients $c_1$, $c_2$, 
and $c_3$. For larger $t_\text{n}$ (long after the end of beam power), 
we have the asymptotic behaviors 
\begin{align}
& h(t_\text{n})  \approx \frac{2\sqrt{t_\text{n} } }{\sqrt{\pi}} + \cdots 
\nonumber \\
& T_\text{nd}(t_\text{n}) \approx \frac{P_\text{d,nd}^{(0)}  t_\text{F,n} }{\sqrt{\pi} \sqrt{t_\text{n} - t_\text{F,n}/2 } } + \cdots \;\;  
\text{ for } t_\text{n} > t_\text{F,n}
\nonumber
\end{align}
We use the asymptotic formula to estimate the settlement time $t_\text{stl,n} $.  
\begin{align} 
& T_\text{nd}(t_\text{stl,n} ) = c_\text{stl,n} \equiv 0.5 
\;\; \Longrightarrow \;\; 
\frac{P_\text{d,nd}^{(0)}  t_\text{F,n} }{\sqrt{\pi} \sqrt{t_\text{n} - t_\text{F,n}/2 } }
 \approx c_\text{stl,n} 
\nonumber \\
& \qquad \Longrightarrow \;\; 
t_\text{stl,n}  \approx \frac{1}{2} t_\text{F,n} + 
\Big( \frac{P_\text{d,nd}^{(0)}  t_\text{F,n} }{c_\text{stl,n} \sqrt{\pi} }  \Big)^2 
\end{align}
where $c_\text{stl,n} $ is the coefficient defining the settlement time. $c_\text{stl,n} = 0.5$ 
gives the time when the maximum temperature falls back half way between 
the skin baseline temperature and the nociceptor activation temperature.

We use Simpson's rule to compute the integral in (\ref{Omega_calc}). 
Since the beam power is turn off at $t_\text{F,n}$, 
this discontinuity in heat source produces a weak singularity in 
the temperature at $t_\text{F,n}$ (the time derivative is discontinuous). 
To minimize the discretization error in numerical integration, 
we set $t_\text{F,n}$ as a grid point. 
We use a uniform grid in the interval $[0, t_\text{F,n}]$ and 
a non-uniform grid (gradually increasing in grid size) in 
the interval $[t_\text{F,n}, t_\text{stl,n} ]$. 
The union of two grids covers $[0, t_\text{stl,n} ]$. 
The integrand function is infinitely differentiable inside each interval. 
Specifically, we set the two grids as 
\begin{align}
& t^{(1)}_j = \frac{j}{N_1} t_\text{F,n}, \quad j = 0, 1, 2, \ldots, N_1 
\label{grid_1} \\ 
& t^{(2)}_j = t_\text{F,n} {\big(1-(1-\beta) \frac{j}{N_2} \big)^{-2 }}, 
\quad \beta =\sqrt{\frac{t_\text{F,n} }{t_\text{stl,n} } }, \quad j = 0, 1, 2, \ldots, N_2
\label{grid_2}
\end{align}
The thermal damage parameter 
$\Omega $ given in (\ref{Omega_calc}) is discretized as 
\begin{align}
& \Omega = \int_0^{t_\text{F,n}} k_\text{d}(t_\text{n}) dt_\text{n} +
\int_{t_\text{F,n}}^{t_\text{stl,n}} k_\text{d}(t_\text{n}) dt_\text{n}
\approx \Omega_1 + \Omega_2  
\label{Omega_calc2} \\
& \quad \Omega_1 = \sum_{j=0}^{N_1-1} \Big( k_\text{d}(t^{(1)}_j) + 4 k_\text{d}(t^{(1)}_{j+\frac{1}{2}} ) 
+ k_\text{d}(t^{(1)}_{j+1}) \Big) \frac{(\Delta t)^{(1)}_j}{6}
\nonumber \\
& \qquad t^{(1)}_{j+\frac{1}{2}} \equiv \frac{1}{2}(t^{(1)}_{j} + t^{(1)}_{j+1}), \qquad
(\Delta t)^{(1)}_j \equiv t^{(1)}_{j+1}-t^{(1)}_{j} 
\nonumber \\
& \quad \Omega_2 = \sum_{j=0}^{N_2-1} \Big( k_\text{d}(t^{(2)}_j) + 4 k_\text{d}(t^{(2)}_{j+\frac{1}{2}} ) 
+ k_\text{d}(t^{(2)}_{j+1}) \Big) \frac{(\Delta t)^{(2)}_j}{6}
\nonumber \\
& \qquad t^{(2)}_{j+\frac{1}{2}} \equiv \frac{1}{2}(t^{(2)}_{j} + t^{(2)}_{j+1}), \qquad
(\Delta t)^{(2)}_j \equiv t^{(2)}_{j+1}-t^{(2)}_{j} 
\nonumber
\end{align}

\subsection{Simulation results based on the listed parameters}
Given the parameter values listed in subsection \ref{para_values}, an exposure 
test in our model is completely specified by $(t_F, r_b/r_s)$. 
In this setting, the absorbed beam center power density 
$P_d^{(0)}$ is hidden in $t_F$; the effect of the unknown volume threshold $v_c$
is hidden in the unknown lateral scale $r_s$. 
This setting is necessary for getting around the unknown $(P_d^{(0)}, v_c)$. 

We first demonstrate how the beam radius $r_b/r_s$ influences $T(t)$, the 
skin surface physical temperature at the beam center as a function of physical time $t$, 
which determines the thermal damage parameter $\Omega$. 
\begin{figure}[!h]
\vskip 0.5cm
\begin{center}
\psfig{figure=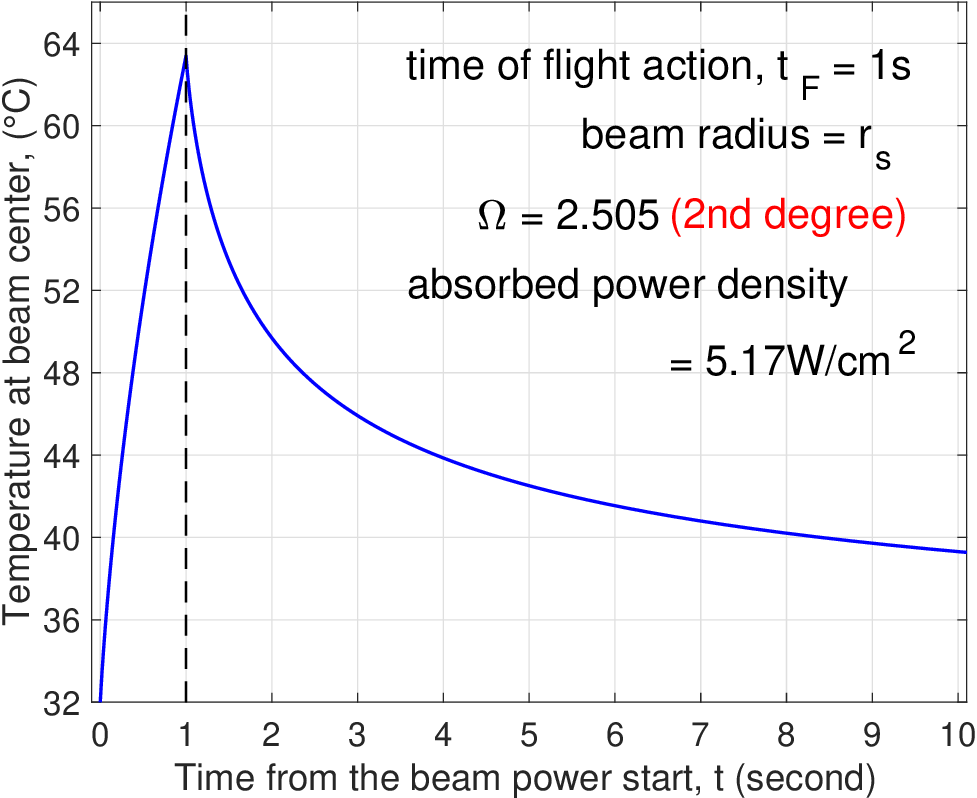, width=2.8in} \quad
\psfig{figure=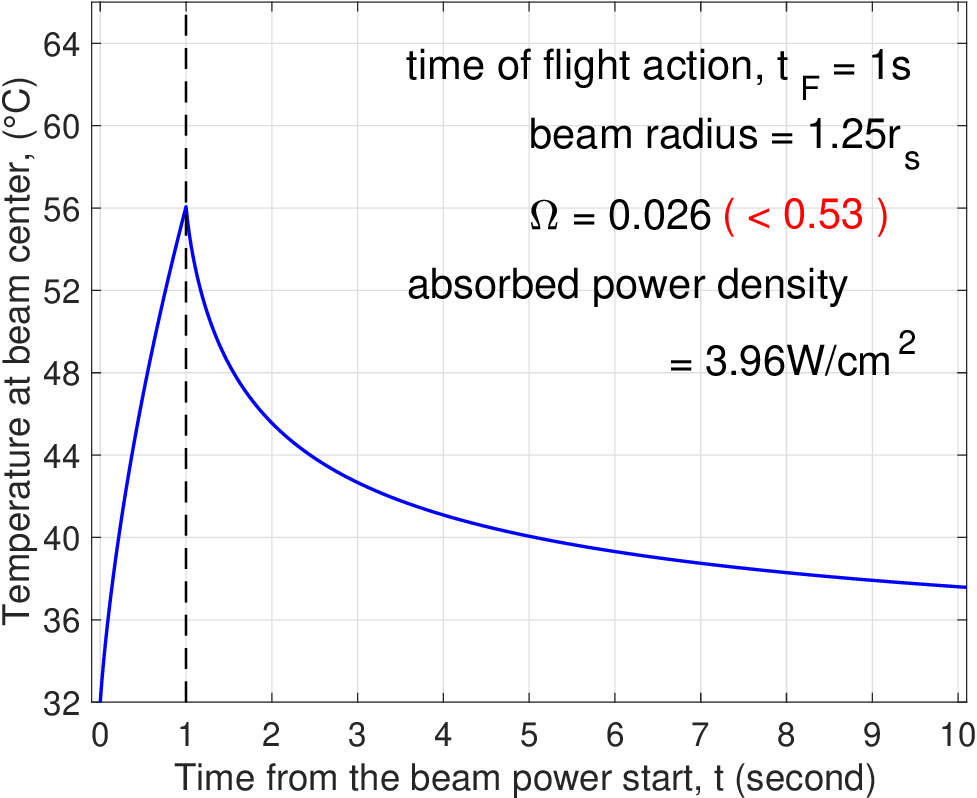, width=2.8in} \\
\hspace*{0.5cm} (a) \hspace*{7cm} (b)
\end{center}
\vskip -0.75cm
\caption{Results for $t_\text{F} = 1\text{s}$. Skin surface temperature vs $t$.
(a) $r_b = r_s$; (b) $r_b = 1.25 r_s$. }
\label{fig_01}
\end{figure}
In Figure \ref{fig_01}, the observed time of flight action is fixed at $t_F = 1 \text{s}$. 
The left panel shows the results of an exposure test with $r_b =r_s$ where $r_s$ 
is the unknown lateral length scale derived from the unknown volume threshold $v_c$. 
The skin surface temperature increases monotonically in $[0, t_F]$, 
driven by the electromagnetic heating. 
After the beam power is turned off at $t_F$, the temperature decreases monotonically
for $t > t_F$. The maximum temperature reached is $T_\text{max} = 63.39 ^\circ\text{C}$. 
The temperature remains high for a substantial period of time after $t_F$, 
which contributes to the thermal damage parameter $\Omega $.
For $t_F=1 \text{s}$, the corresponding absorbed beam power density is 
$P_d^{(0)} = 5.17 \text{W}/\text{cm}^2$ and the thermal damage parameter is
$\Omega = 2.505$, which indicates a second-degree burn ($\Omega > 1$). 
The right panel of Figure \ref{fig_01} shows the results for a slightly larger 
beam radius $r_b =1.25 r_s$. 
The maximum temperature of the time course drops to 
$T_\text{max} = 56.06 ^\circ\text{C}$; 
the absorbed beam power density drops to $P_d^{(0)} = 3.96 \text{W}/\text{cm}^2$; and 
the thermal damage parameter drops to $\Omega = 0.026$, which is far below 
the threshold of the first-degree burn ($\Omega > 0.53$). 
Comparing the two panels of Figure \ref{fig_01}, we see that the same flight action time 
of $t_F=1\text{s}$ can be achieved with a much lower thermal injury risk if we use a 
larger beam radius with a lower beam power density. 
A larger beam radius heats a larger skin area and reduces the activated depth 
needed for reaching the volume threshold. 
Here the activated depth refers to the depth of the activated 3D skin region. 
Since the electromagnetic heating source decays exponentially in depth, 
a smaller activated depth requires a lower beam power density and produces a 
lower maximum temperature, which in turn lowers 
the thermal damage parameter $\Omega$.  

\begin{figure}[!h]
\vskip 0.5cm
\begin{center}
\psfig{figure=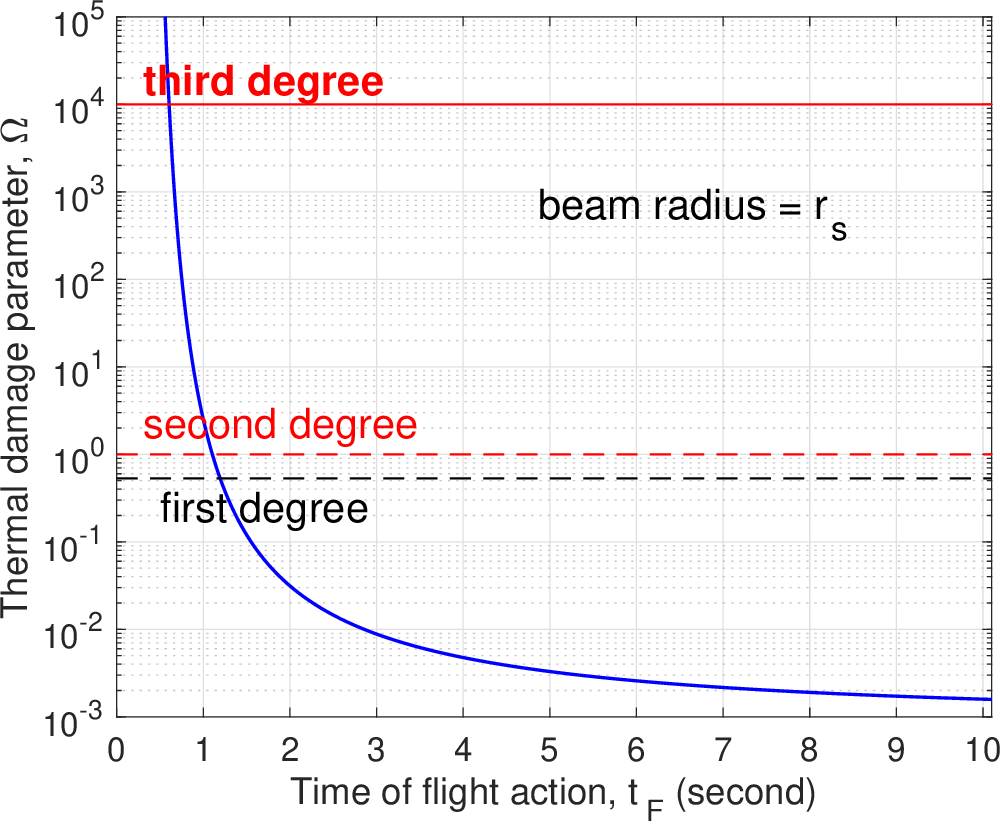, width=2.8in} \quad
\psfig{figure=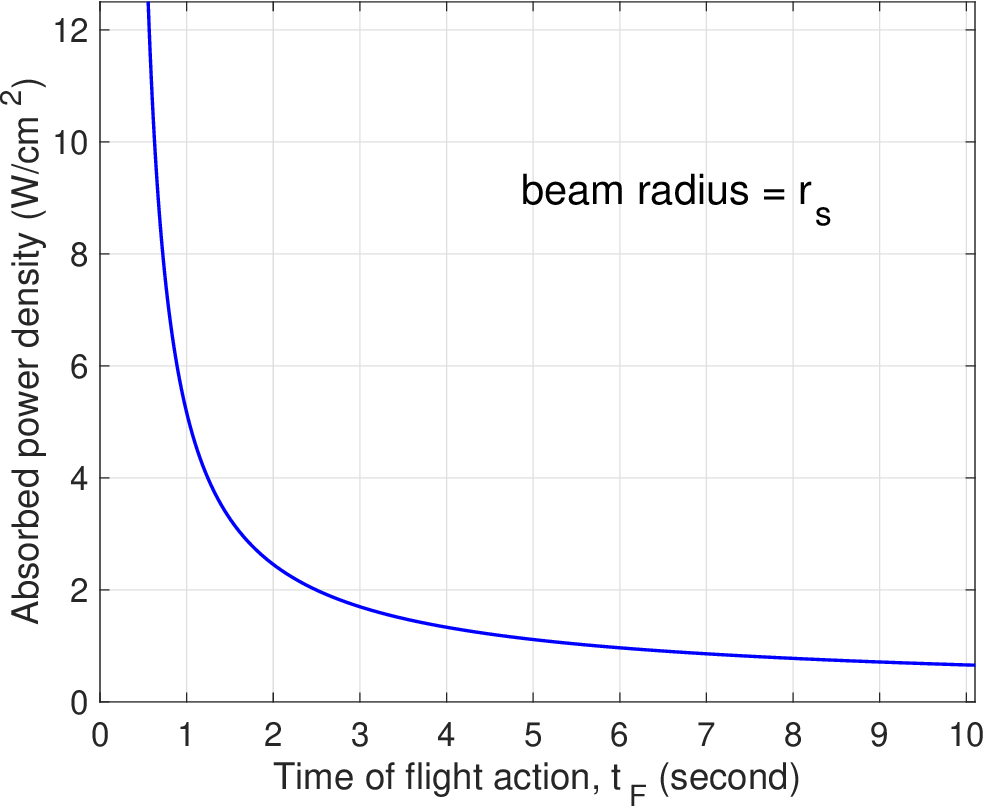, width=2.8in} \\
\hspace*{0.5cm} (a) \hspace*{7cm} (b)
\end{center}
\vskip -0.75cm
\caption{Results for $r_b =r_s$. (a) the thermal damage parameter $\Omega$ vs 
the observed flight action time $t_\text{F}$; 
(b) the \underline{absorbed beam power density} $P_d^{(0)}$ vs $t_\text{F}$.} 
\label{fig_02}
\end{figure}

In Figure \ref{fig_02}, we fix the beam radius at $r_b = r_s$. We examine how 
the thermal damage parameter $\Omega $ and the absorbed beam power density $P_d^{(0)}$
vary with the observed time of flight action $t_F$. 
It is important to point out that in exposure tests, the time of flight action $t_F$
is not precisely prescribed/controlled in the experiment setup; 
the time of flight action $t_F$ is naturally \underline{observed} reflecting 
the absorbed beam power density $P_d^{(0)}$  that is actually materialized 
in each test, which is not measurable. 
The experimenter may adjust the power density at the beam source to influence 
$P_d^{(0)}$ (and thus, to influence $t_F$).  
As the observed $t_F$ increases, both $\Omega$ and $P_d^{(0)}$ decrease 
monotonically. 
At a fixed beam radius, the activated depth needed for reaching the volume threshold 
is also fixed. However, for a larger observed $t_F$ (and thus a larger $t_c$), 
the heat conduction has more time to smooth out the temperature distribution 
along the depth. A more uniform temperature along the depth means that 
the same activated depth is achieved with a lower maximum temperature, 
which in turn lowers the thermal damage parameter $\Omega$. 

\begin{figure}[!h]
\vskip 0.5cm
\begin{center}
\psfig{figure=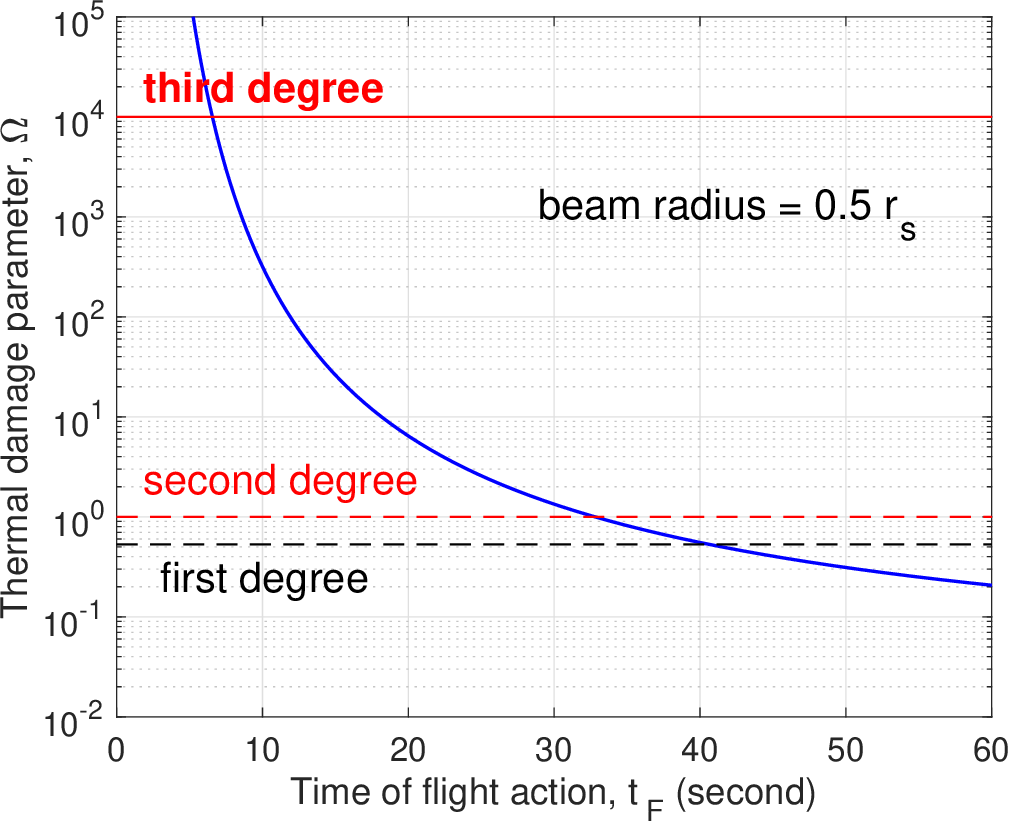, width=2.8in} \quad
\psfig{figure=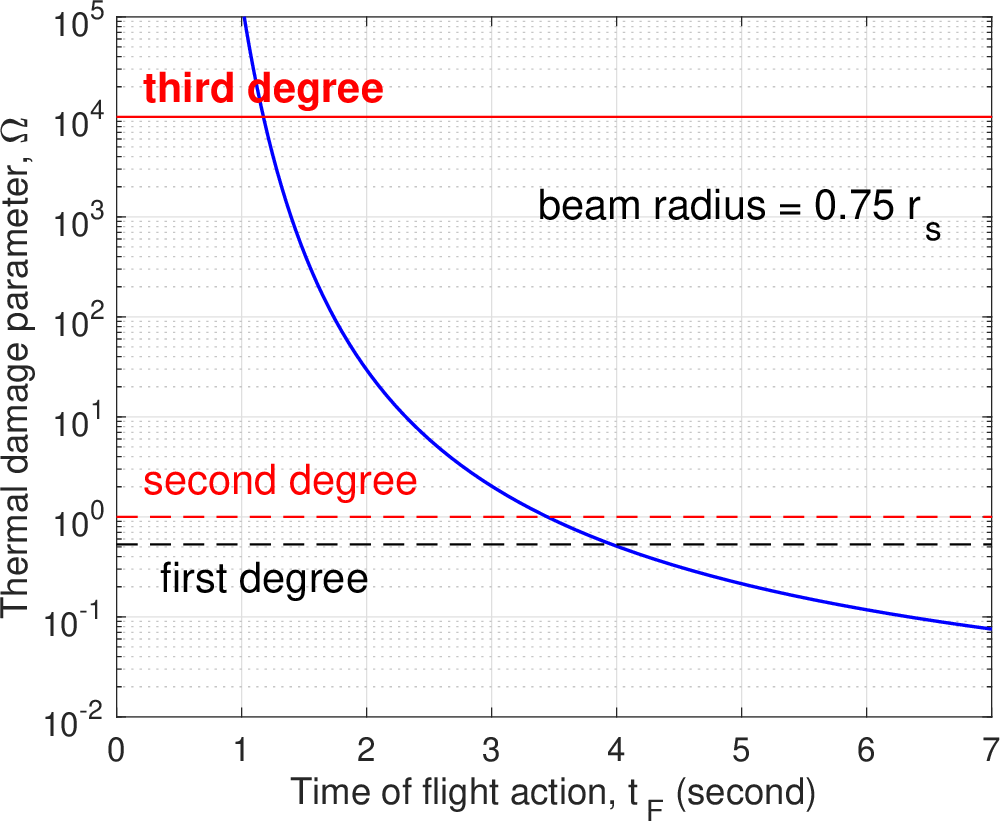, width=2.8in} \\[2ex]
\psfig{figure=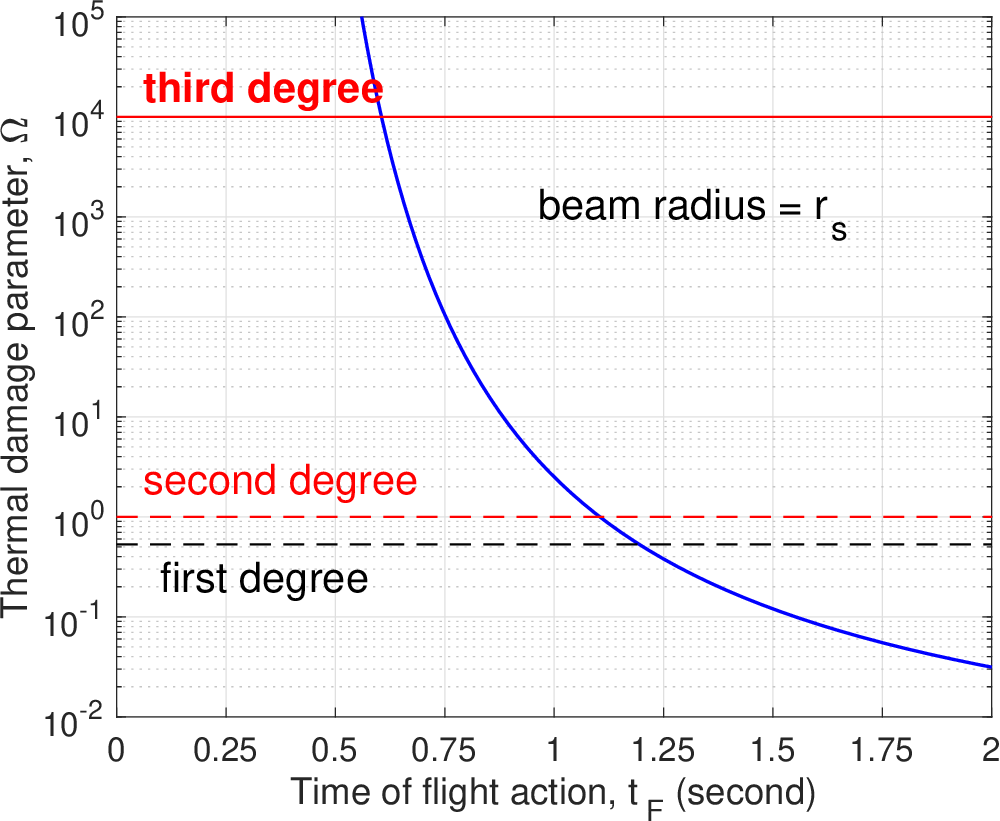, width=2.8in} \quad
\psfig{figure=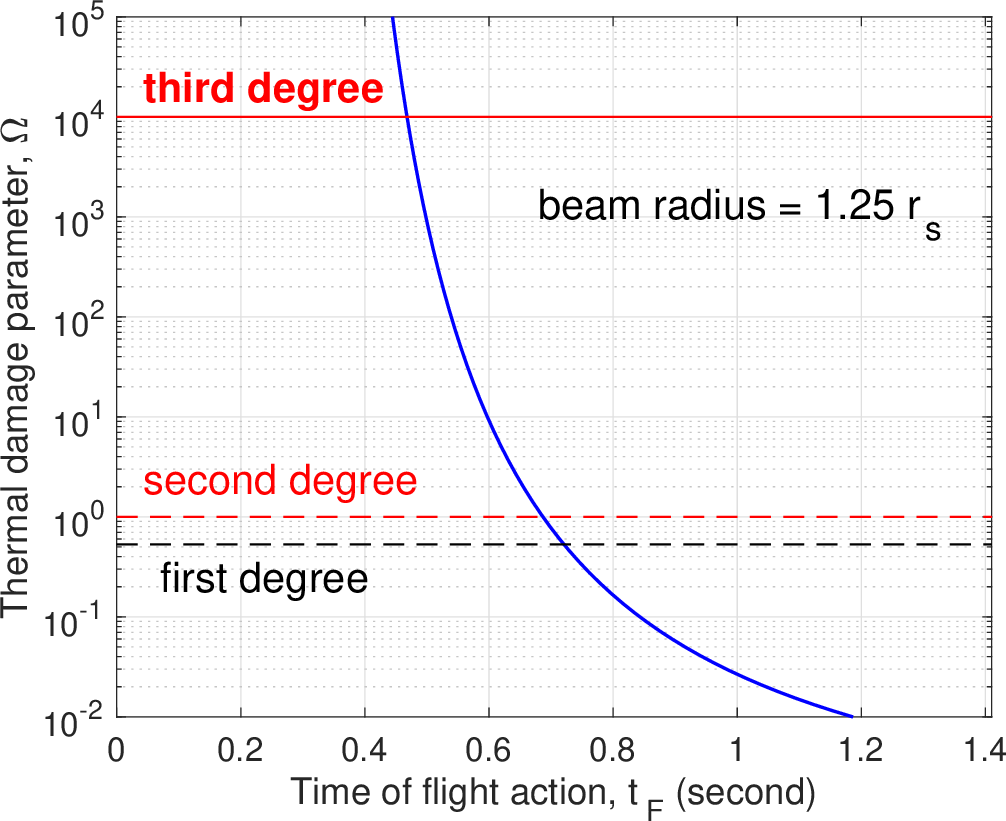, width=2.8in}
\end{center}
\vskip -0.75cm
\caption{$\Omega$ vs $t_\text{F}$ at $r_b = 0.5r_s$, $0.75r_s$, $ r_s$, and $1.25 r_s$.} 
\label{fig_03}
\end{figure}

\begin{figure}[!h]
\vskip 0.5cm
\begin{center}
\psfig{figure=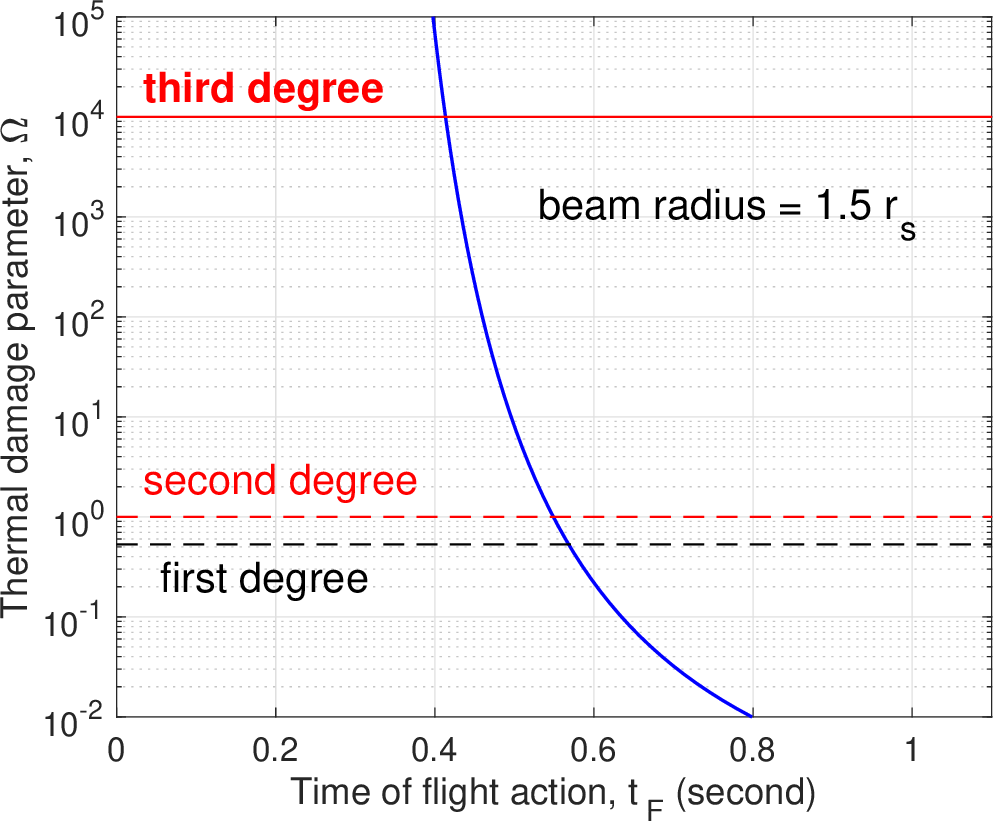, width=2.8in} \quad
\psfig{figure=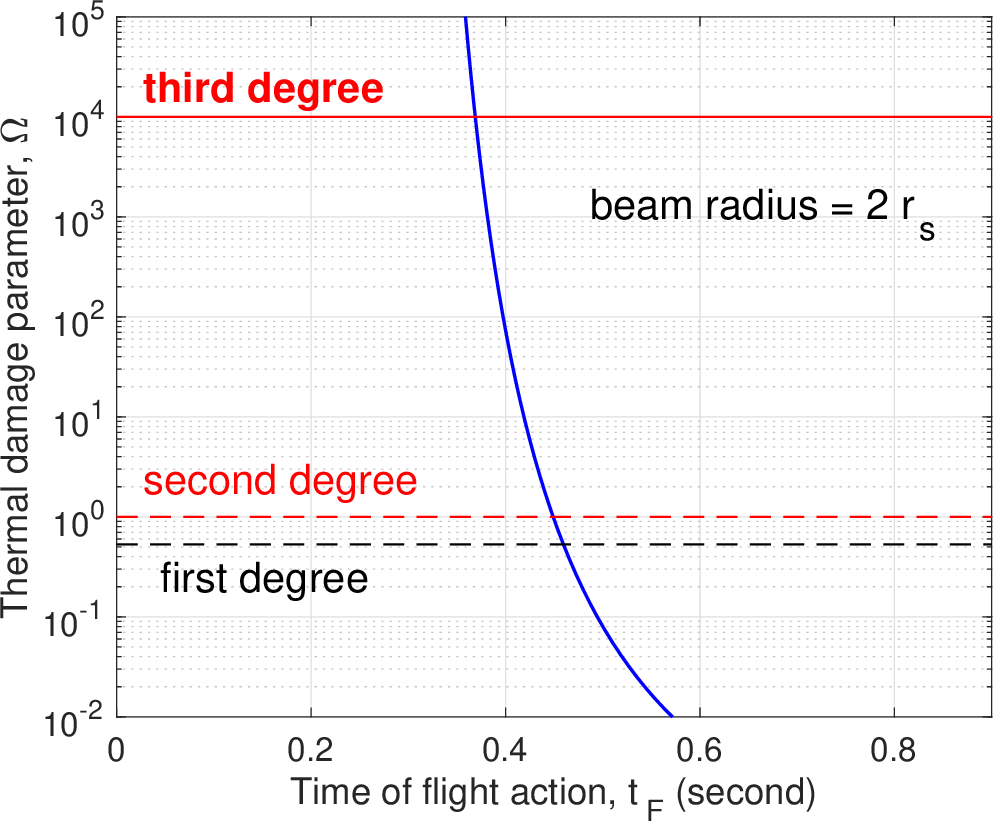, width=2.8in} \\[2ex]
\psfig{figure=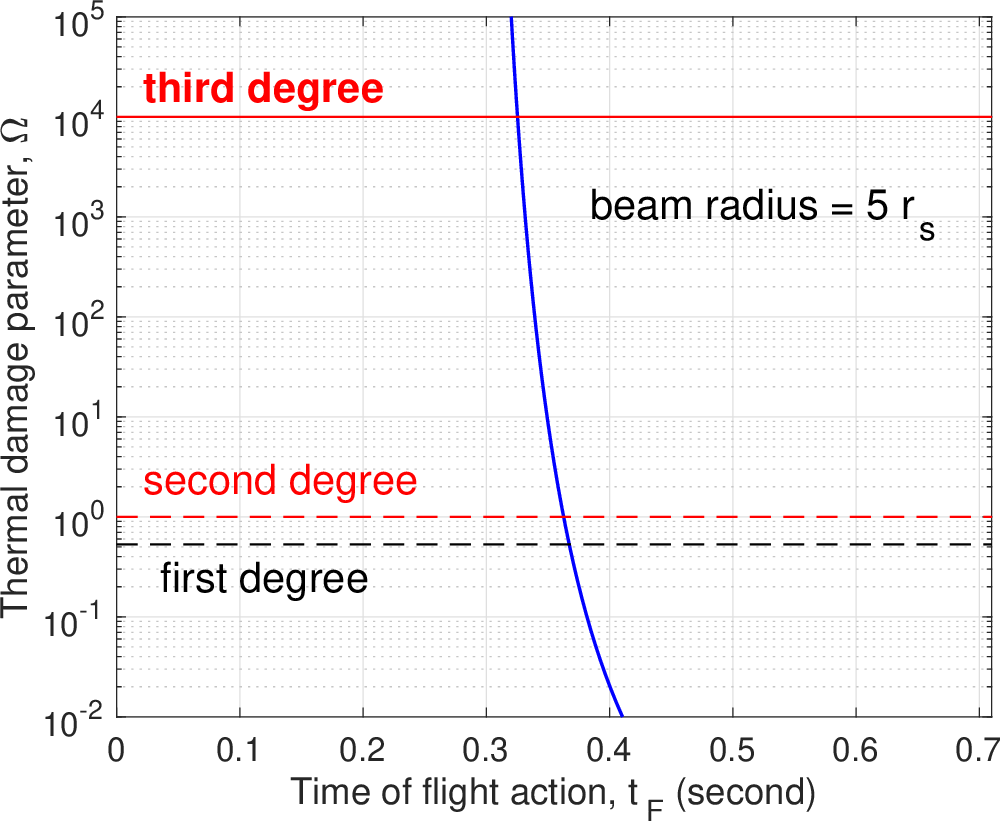, width=2.8in} \quad
\psfig{figure=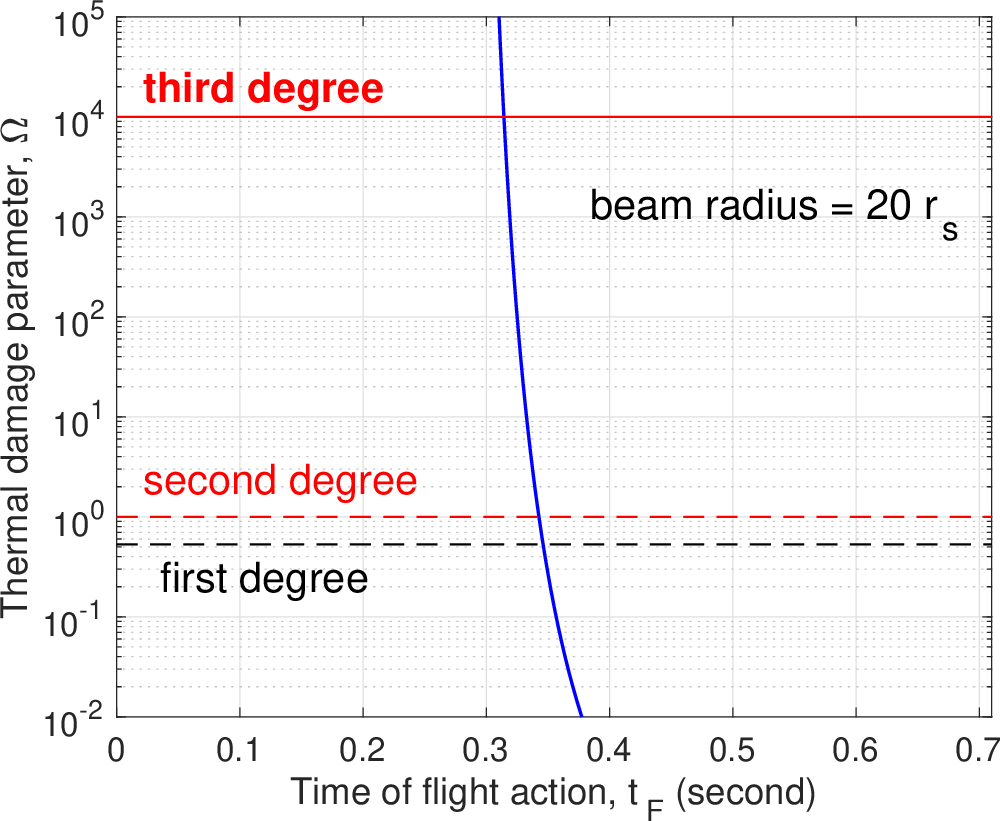, width=2.8in}
\end{center}
\vskip -0.75cm
\caption{$\Omega$ vs $t_\text{F}$ at $r_b = 1.5r_s$, $2 r_s$, $5 r_s$, and $20 r_s$.} 
\label{fig_04}
\end{figure}

Next we explore the effect of beam radius $r_b$ on the relation of $\Omega $ vs $t_F$. 
Figure \ref{fig_03} and Figure \ref{fig_04} show results for 8 values of $r_b$ ranging
from $r_b = 0.5 r_s$ to $r_b = 20 r_s$. 
Similar to what was demonstrated in Figure \ref{fig_02},
for all values of $r_b$, the thermal damage parameter $\Omega$ is always 
a decreasing function of $t_F$. In all panels, the vertical axis for $\Omega $
has the same logarithmic scale with thresholds labeled for 1st, 2nd and 3rd-degree burns. 
The horizontal axis for $t_F$ is set differently for each individual $r_b$ to capture 
the transition of $\Omega $ from no injury (below the 1st-degree burn) to 
the 2nd-degree burn. 
As the beam radius $r_b$ increases, the transition window in $t_F$ decreases rapidly. 
For $r_b=0.5r_s$, the transition window spans $t_F = 33 \sim 40\text{s}$ 
(the top left panel of Figure \ref{fig_03}). 
With a beam of radius $r_b=0.5r_s$, it is impossible to make flight action occur  
below 30 seconds while avoiding thermal injury at the same time. 
When the beam radius is increased to $r_b=1.25r_s$, the transition window drops to 
$t_F \approx 0.7\text{s}$ (the bottom right panel of Figure \ref{fig_03}). 
When the beam radius is further increased to $r_b=5r_s$, the transition window 
drops further to  
$t_F \approx 0.37\text{s}$ (the bottom left panel of Figure \ref{fig_04}). 
It is clear that if we want to flight action to occur at a small $t_F$ 
while avoiding thermal injury, a large beam radius should be used.

There is, however, a lower bound on $t_F$. 
Since $t_F = t_c + t_R$ and $t_R = 0.275\text{s}$ is independent of the beam 
specifications, the smallest achievable $t_F$ is definitely bounded from below 
by $0.275\text{s}$. 
This obvious lower bound, however, is not attainable when we add in 
the requirement of avoiding thermal injury. To make $t_F \approx t_R$, 
we need to push the flight initiation time $t_c$ to zero, 
which requires a very large beam power density to heat skin from 
its baseline temperature ($T_\text{base}$) to the nociceptor activation temperature 
($T_\text{act}$) almost instantly. 
This very large beam power density does not end at the flight initiation $t_c$. 
It continues until the flight action at $t_F = t_c + t_R$. 
This very large beam power density over a time period of $t_F > 0.275\text{s}$ 
definitely will cause a significant thermal injury. 
The practical lower bound for achievable $t_F$ that avoids thermal injury is about 
$t_F \approx 0.347\text{s}$ at a very large beam radius of $r_b = 20r_s$ 
(the bottom right panel of Figure \ref{fig_04}). 

\begin{figure}[!h]
\vskip 0.5cm
\begin{center}
\psfig{figure=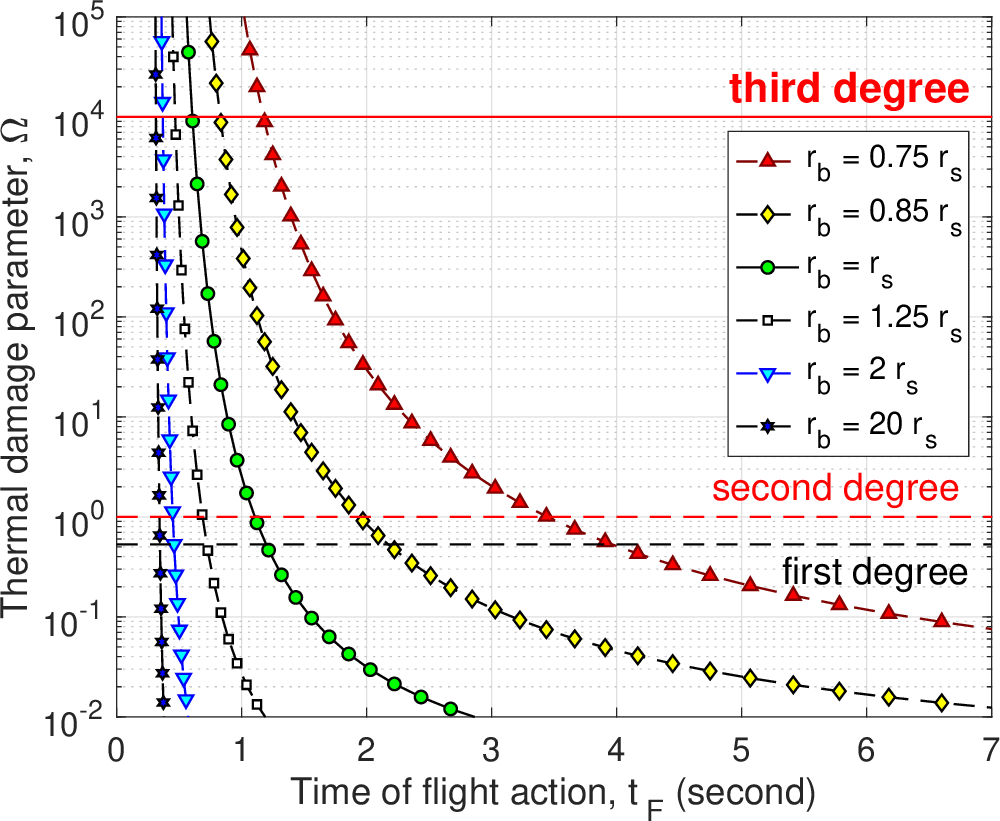, width=4.0in} 
\end{center}
\vskip -0.75cm
\caption{Curves of $\Omega$ vs $t_F$ for several values of beam radius $r_b$.} 
\label{fig_05}
\end{figure}

In Figure \ref{fig_05}, we compare six curves of $\Omega$ vs $t_F$ 
for beam radius ranging from $r_b=0.75r_s$ to $r_b=20r_s$, plotted in one panel. 
As the beam size $r_b$ is increased from slightly below $r_s$ to $r_s$ and 
to slightly above $r_s$, the location of the transition window, 
from no injury to the 2nd-degree burn, drops very rapidly in $t_F$. 
When $r_b =0.75r_s$, $t_F = 3\text{s}$ is definitely not safe for avoiding thermal injury.
When $r_b =1.25r_s$, it is safe even at $t_F = 1\text{s}$. As $r_b$ is increased further above
$r_s$, the location of the transition window converges to a 
practical lower bound of about $t_F = 0.347\text{s}$. 
This attainable lower bound is noticeably 
about the human reaction time $t_R=0.275\text{s}$. As we discussed above, the attainable 
lower bound on $t_F $ not only is limited by $t_R$ but is also constrained by 
no thermal injury. 
\subsection{Sensitivity study}
We study the sensitivity of $\Omega$ vs $t_F$ when the parameters 
are perturbed from their listed values in subsection \ref{para_values}.
In our sensitivity study, we vary one parameter at a time while fixing others 
at the listed values. 

Recall that in our model formulation, the beam used in an exposure test is completely 
described by $(t_F, r_b/r_s)$ where the absorbed beam power density 
$P_d^{(0)}$ is hidden in the observed time of flight action $t_F$. 
The beam radius is specified as a multiple of the lateral length scale 
$r_s \equiv \sqrt{\frac{\mu v_c}{\pi}} $, which varies with parameters 
$\mu$ and $v_c$ but is independent of other parameters. 
As a result, the sensitivity study with respect to $\mu$ or $v_c$ requires 
additional adjustment when setting dimensionless parameters. 
We first study the sensitivity with respect to parameters that do not affect 
the lateral length scale $r_s$. We fix the beam radius at $r_b = r_s$, 
which means the physical beam radius is fixed as the parameter is perturbed. 

\begin{figure}[!h]
\vskip 0.5cm
\begin{center}
\psfig{figure=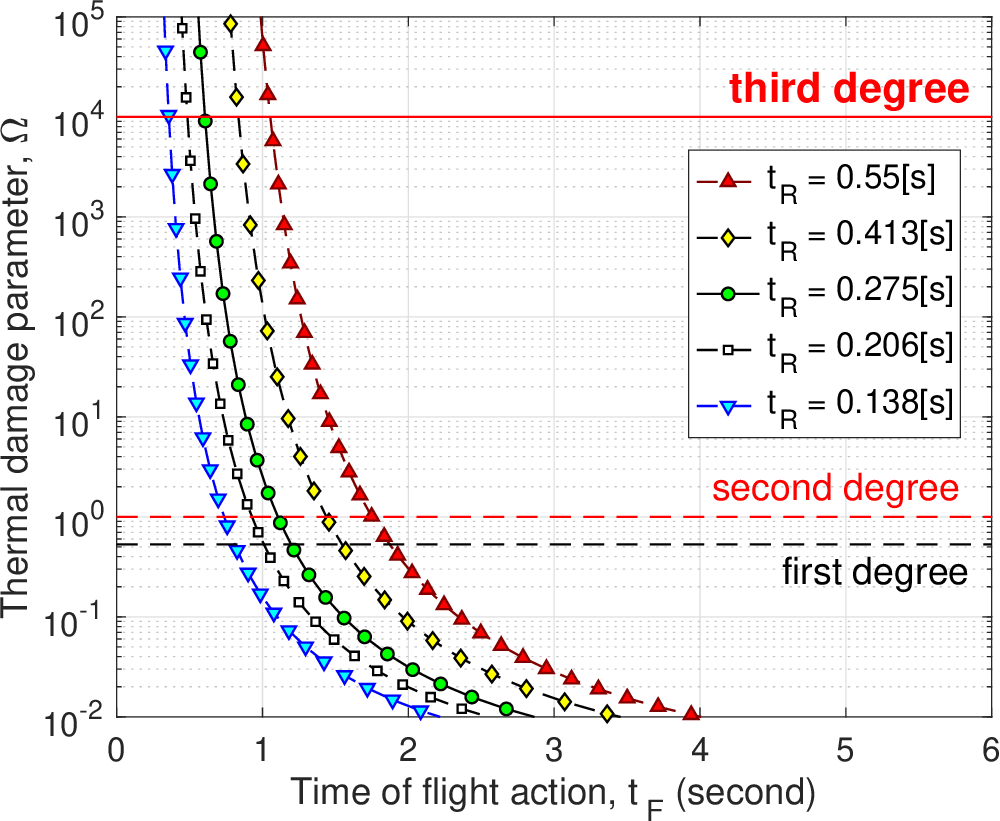, width=4.0in} 
\end{center}
\vskip -0.75cm
\caption{Curves of $\Omega$ vs $t_F$ for several values of human reaction time $t_R$. 
$r_b = r_s$.}
\label{fig_07}
\end{figure}
\begin{figure}[!h]
\vskip 0.5cm
\begin{center}
\psfig{figure=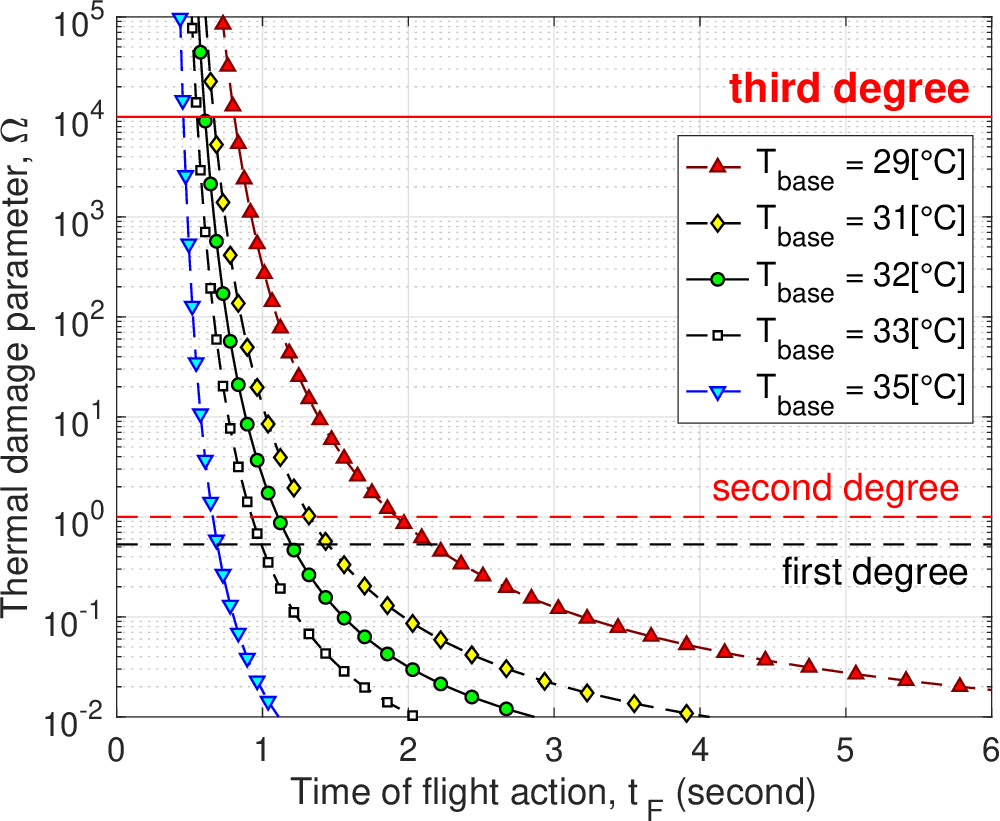, width=4in} 
\end{center}
\vskip -0.75cm
\caption{Curves of $\Omega$ vs $t_F$ for several values of skin baseline temperature $T_{base}$.}
\label{fig_08}
\end{figure}
We start with the human reaction time $t_R$. 
Curves of $\Omega$ vs $t_F$ for values of $t_R$ ranging from 0.138s to 0.55s
are plotted in Figure \ref{fig_07}. 
The listed value of $t_R$ in subsection \ref{para_values} is $t_R =0.275\text{s}$.
As shown in Figure \ref{fig_07}, at any given time of flight action $t_F$, 
the thermal damage parameter $\Omega$ increases monotonically with 
the human reaction time $t_R$. This result is intuitive. At a fixed $t_F=t_c+t_R$, 
a larger $t_R$ gives a smaller time to flight initiation $t_c$, 
which requires a higher beam power density for a more rapid heating. 
A higher heating power over a fixed heating duration $t_F$
increases the thermal damage risk.

We explore the effect of the skin baseline temperature $T_\text{base}$. 
The listed value of $T_\text{base}$ in subsection \ref{para_values} is 
$T_\text{base} =32 ^\circ\text{C}$. 
Curves of $\Omega$ vs $t_F$ for values of $T_\text{base}$ 
ranging from $29 ^\circ\text{C}$ to $35 ^\circ\text{C}$ are plotted 
in Figure \ref{fig_08}, which shows that at any given time of flight action 
$t_F$, the thermal damage parameter $\Omega$ decreases 
if the skin baseline temperature $T_\text{base}$ is increased. 
At the first glance, this results is not immediately intuitive. 
It is explained by the effect of $T_\text{base}$ on the power density needed. 
A higher baseline temperature $T_\text{base}$ implies that  
a smaller temperature increase is needed for reaching the nociceptor activation temperature, 
At a fixed $t_c = t_F - t_R$, a smaller temperature increase translates to 
a lower beam power density, which in turn decreases the thermal damage risk.
\begin{figure}[!h]
\vskip 0.5cm
\begin{center}
\psfig{figure=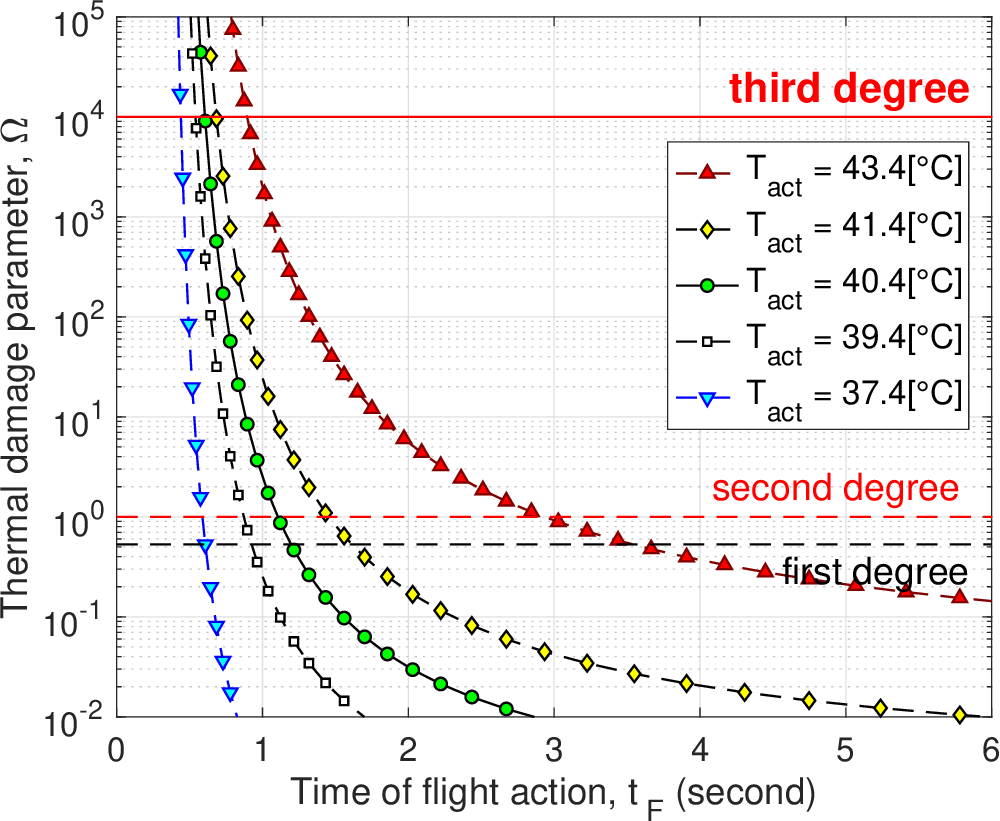, width=4in} 
\end{center}
\vskip -0.75cm
\caption{Curves of $\Omega$ vs $t_F$ for several values of 
nociceptor activation temperature $T_{act}$.} 
\label{fig_09}
\end{figure}

We continue onto the effect of the nociceptor activation temperature $T_\text{act}$. 
The listed value of $T_\text{act}$ in subsection \ref{para_values} is 
$T_\text{act} =40.4 ^\circ\text{C}$.
Figure \ref{fig_09} shows curves of $\Omega$ vs $t_F$ for 
values of $T_\text{act}$ ranging from $37.4 ^\circ\text{C}$ to $43.4 ^\circ\text{C}$. 
At any given time of flight action $t_F$, 
the thermal damage parameter $\Omega$ increases monotonically with
the nociceptor activation temperature $T_\text{act}$. 
Similar to the situation with $T_\text{base}$, the result of $T_\text{act}$
is explained by its effect on the power density needed. 
A higher $T_\text{act}$ means a larger temperature increase is needed to reach  
$T_\text{act}$ from $T_\text{base}$, which translates to a higher beam power density
and thus a higher thermal damage risk. 
\begin{figure}[!h]
\vskip 0.5cm
\begin{center}
\psfig{figure=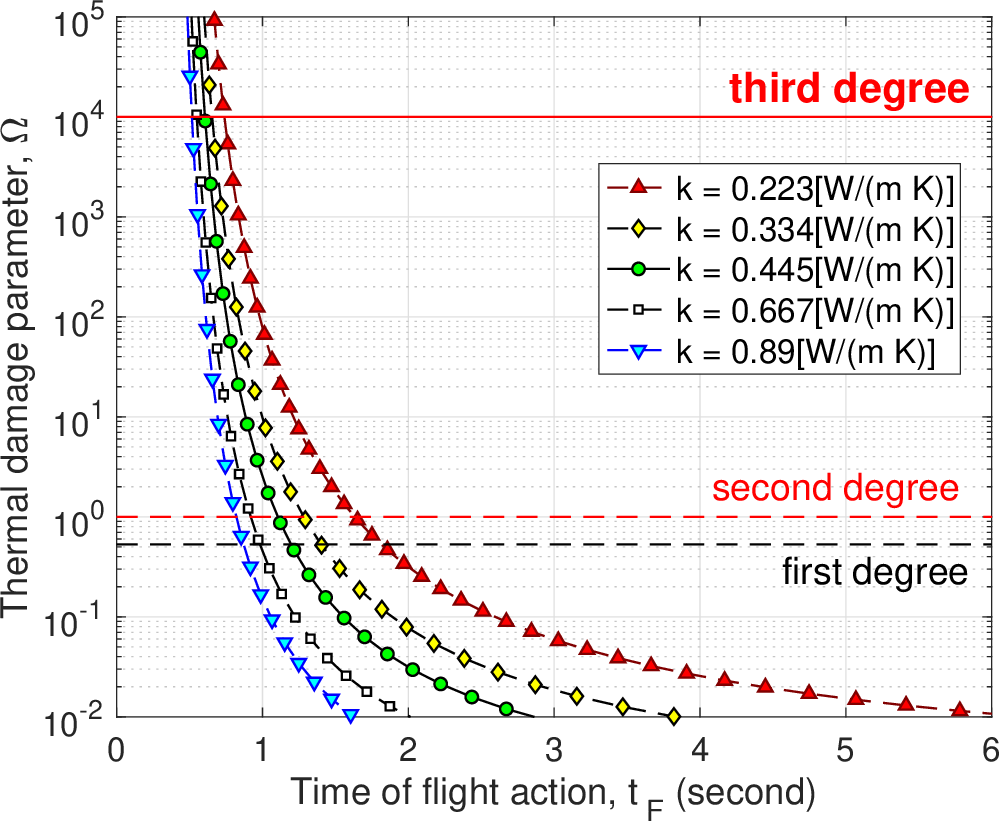, width=4in} 
\end{center}
\vskip -0.75cm
\caption{Curves of $\Omega$ vs $t_F$ for several values of 
skin heat conductivity $k$.} 
\label{fig_10}
\end{figure}

Next we investigate the effect of the skin heat conductivity $k$. 
The listed value of $k$ in subsection \ref{para_values} is $k =0.445 \text{W/(mK)}$.
We plots curves of $\Omega$ vs $t_F$ for values of $k$ 
ranging from $0.223 \text{W/(mK)}$ to $0.89 \text{W/(mK)}$ 
in Figure \ref{fig_10}. 
At any given time of flight action $t_F$, 
the thermal damage parameter $\Omega$ decreases as the skin heat conductivity
$k$ is increased. We need to make sense of this result in our model framework. 
At a fixed $t_c = t_F-t_R$, a larger $k$ smoothes out the temperature more in depth 
via heat conduction and makes the temperature increase in depth more uniform. 
A more uniform heating lowers the beam power density required for reaching 
the activated depth and lowers the maximum skin temperature, 
which decreases the thermal damage risk. 
\begin{figure}[!h]
\vskip 0.5cm
\begin{center}
\psfig{figure=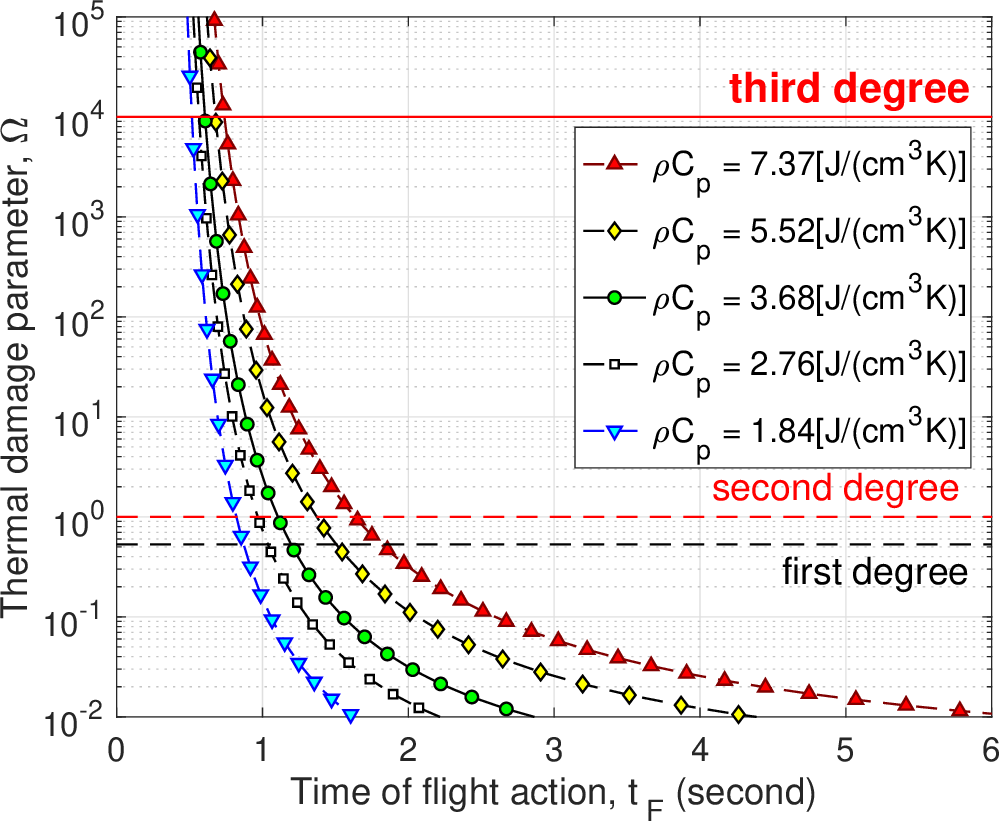, width=4in} 
\end{center}
\vskip -0.75cm
\caption{Curves of $\Omega$ vs $t_F$ for several values of 
volumetric heat capacity $\rho C_p$.} 
\label{fig_11}
\end{figure}

Another quantity that influences the effectiveness of heat conduction in smoothing 
out the temperature is the volumetric heat capacity $\rho C_p$. 
In our model, parameters $\rho $ and $C_p$ appear only in the combination $\rho C_p$. 
The listed value of $\rho C_p$ in subsection \ref{para_values} is 
$\rho C_p=3.68\times 10^6 \text{J/(m$^3$K)}=3.68 \text{J/(cm$^3$K)}$.
Figure \ref{fig_11} shows curves of $\Omega$ vs $t_F$ for 
values of $\rho C_p$ ranging from $1.84 \text{J/(cm$^3$K)}$ to 
$7.37 \text{J/(cm$^3$K)}$. 
At any given time of flight action $t_F$, 
the thermal damage parameter $\Omega$ increases monotonically with
the volumetric heat capacity $\rho C_p$. 
The mechanism behind this result is similar to that for parameter $k$ in Figure \ref{fig_10}. 
At a fixed conductivity $k$, a larger volumetric heat capacity $\rho C_p$ 
renders the heat conduction less effective in smoothing out the temperature in depth 
because more heat needs to be 
transferred to reduce a given temperature gradient. Thus, a larger 
volumetric heat capacity $\rho C_p$ has an effect similar to that of a smaller 
conductivity $k$. 
\begin{figure}[!h]
\vskip 0.5cm
\begin{center}
\psfig{figure=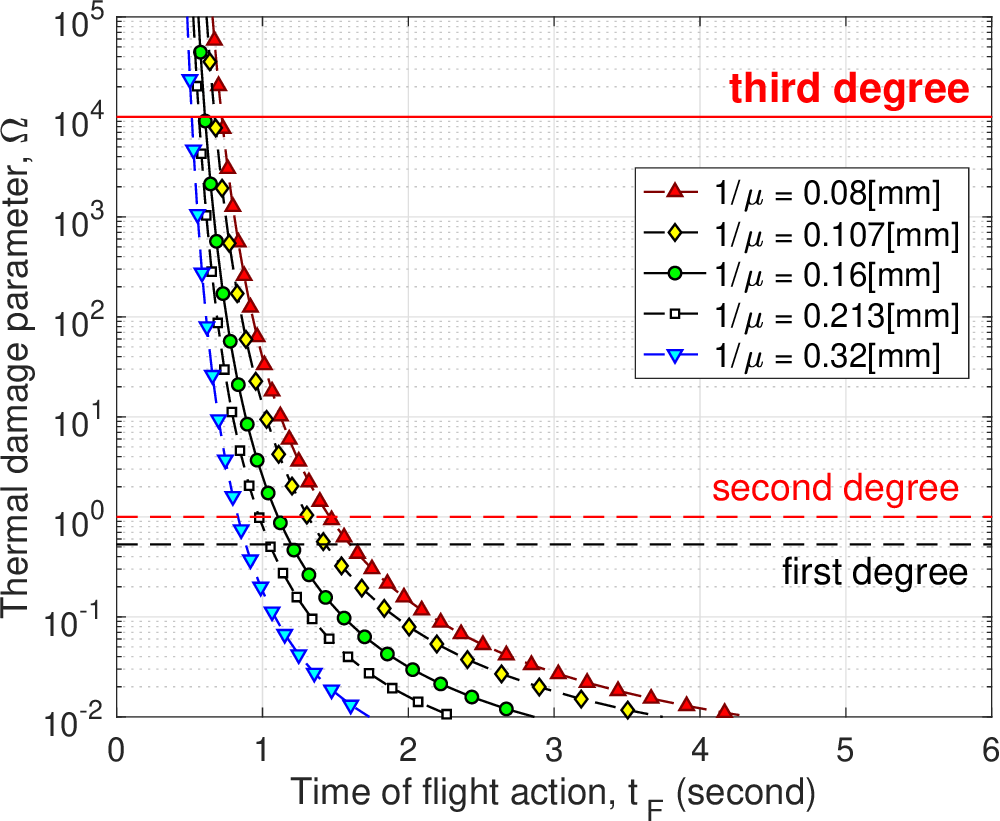, width=4in} 
\end{center}
\vskip -0.75cm
\caption{Curves of $\Omega$ vs $t_F$ for several values of 
penetration depth $1/\mu$. }
\label{fig_06}
\end{figure}

Finally, we examine the effect of the electromagnetic penetration depth $1/\mu$
and that of the volume threshold $v_c$. 
In the model formulation, the beam radius is specified as a multiple of 
the lateral length scale $r_s \equiv \sqrt{\frac{\mu v_c}{\pi}} $, 
which depends on parameters $\mu$ and $v_c$. 
In the sensitivity study, to pinpoint the effect of $\mu$, we need to fix all other 
parameters, including the physical beam radius. 
Operationally, while we vary $\mu$ in simulations, it is desirable to continue 
working with the normalized formulation described in subsection \ref{norm_form}.
To accommodate both of these, we use the listed value $1/\mu^{(0)} = 0.16\text{mm}$
to define a reference lateral length scale 
$\displaystyle r_s^{(0)} \equiv \sqrt{\frac{\mu^{(0)} v_c}{\pi}} $, 
which is not affected when $\mu$ is varied in the sensitivity study. 
The reference scale $r_s^{(0)}$ is solely for the purpose of specifying 
 $r_b$. In the sensitivity study, we fix the physical beam radius at $r_b = r_s^{(0)}$. 
In simulations using the normalized formulation, as $\mu $ is varied, 
the normalized beam radius $r_{b,n} $ is adjusted to reflect the same 
underlying physical beam radius. 
\begin{equation}
r_{b,n} \equiv \frac{r_b}{r_s} = \frac{r_s^{(0)}}{r_s} \frac{r_b}{r_s^{(0)}}
= \sqrt{\frac{\mu^{(0)}}{\mu}}  
\label{r_bn_adjust} 
\end{equation}
In (\ref{r_bn_adjust}), for a fixed physical beam radius, when $1/\mu$ is increased 
by a factor of 2, the dimensionless $r_{b,n}$ increases by a factor of $\sqrt{2}$. 
In addition, $\mu$ also affects the time scale 
$t_s = \frac{\rho_\text{m} C_p}{k \mu^2}$, power density scale 
$P_\text{s} \equiv k \mu (T_\text{act} -T_\text{base}) $
and the coefficient $c_1 = \frac{\rho_\text{m} C_p}{k \mu^2} A $
for evaluating $\Omega $ in (\ref{Omega_calc}). 
These need to be taken care of properly 
when carrying out simulations in the fully nondimensionalized system. 
Curves of $\Omega$ vs $t_F$ for 
the electromagnetic penetration depth $1/\mu$ 
ranging from 0.08mm to 0.32mm are plotted in Figure \ref{fig_06}. 
The listed value in subsection \ref{para_values} is $1/\mu = 0.16 \text{mm}$.
As shown in Figure \ref{fig_06}, at any given time of flight action $t_F$, 
the thermal damage parameter $\Omega$ decreases as $1/\mu$ is increased. 
A larger $1/\mu$ implies a more uniform heat source in the depth. 
When heating the skin to a given activated depth, a more uniform heating 
leads to a more uniform temperature and a lower maximum skin 
temperature, which decreases the thermal damage parameter $\Omega $. 
\begin{figure}[!h]
\vskip 0.5cm
\begin{center}
\psfig{figure=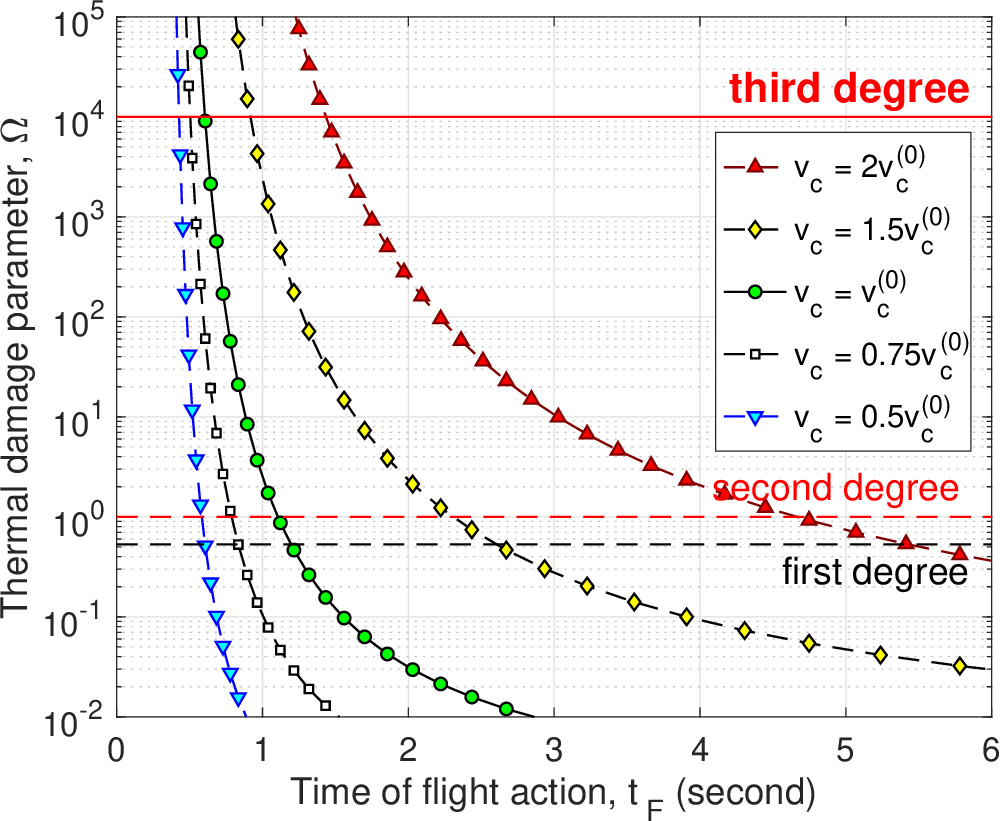, width=4in} 
\end{center}
\vskip -0.75cm
\caption{Curves of $\Omega$ vs $t_F$ for several values of 
volume threshold $v_c$.} 
\label{fig_12}
\end{figure}

In the sensitivity study with respect to the volume threshold $v_c$, we use a similar 
approach to fix the physical beam radius. We select a reference 
$v_c^{(0)} $ to define a reference lateral length scale 
$\displaystyle r_s^{(0)} \equiv \sqrt{\frac{\mu v_c^{(0)}}{\pi}} $. 
Since $v_c$ is unknown, the reference $v_c^{(0)} $ may be arbitrary. 
We investigate the effect of varying $v_c$ relative to this reference value. 
Again, the reference length scale $r_s^{(0)}$ is solely for the purpose of specifying 
the physical beam radius $r_b$. In the sensitivity study, we fix 
the physical beam radius at $r_b = r_s^{(0)}$. 
In simulations using the normalized formulation, as $v_c $ is varied, 
the normalized beam radius $r_{b,n} $ is adjusted to reflect the same underlying 
physical beam radius. 
\begin{equation}
r_{b,n} \equiv \frac{r_b}{r_s} 
= \frac{r_s^{(0)}}{r_s} \frac{r_b}{r_s^{(0)}}
= \sqrt{\frac{v_c^{(0)}}{v_c}}  
\label{r_bn_adjust_2} 
\end{equation}
In (\ref{r_bn_adjust_2}), for a fixed physical beam radius, when $v_c$ is increased 
by a factor of 2, the dimensionless $r_{b,n}$ decreases by a factor of $\sqrt{2}$. 
The most prominent difference between the effects of 
$\mu$ and $v_c$ is that $v_c$ affects only the lateral length scale $r_s$, not other 
parameters. It follows that the effect of increasing $v_c$ while fixing 
the physical beam size $r_b$ is the same as that of reducing $r_b$ while fixing $v_c$. 
Figure \ref{fig_12} shows curves of $\Omega$ vs $t_F$ for 
the volume threshold $v_c$ ranging from $0.5 v_c^{(0)}$ to $2 v_c^{(0)}$ 
where $v_c^{(0)}$ is the reference value. 
At any given time of flight action $t_F$, 
the thermal damage parameter $\Omega$ increases monotonically with $v_c$. 
The trend of $\Omega$ increasing with $v_c$ in Figure \ref{fig_12}
corresponds to the trend of $\Omega $ decreasing with $r_s$ 
observed earlier in Figure \ref{fig_05}. 
%

%%%
%%%% Conclusions
%%%
\section{Conclusions}
In this study, we formulated a model for assessing the skin thermal damage risk 
in exposure tests in which the electromagnetic beam radiation continues until the 
occurrence of flight action when the test subject moves away from the beam 
and/or turns off the beam power. 
The skin temperature increase is driven by the absorbed electromagnetic power
 and is governed by an initial boundary value problem of the heat equation. 
Wherever the local temperature is above the activation threshold, 
thermal nociceptors in skin are activated and transduce an electrical signal. 
The burning sensation is proportional to the aggregated electrical signal
at the exposed skin spot. 
Eventually when the activated skin volume reaches 
a volume threshold, the nociceptor signal is strong enough and the flight is initiated. 
It takes time for the nociceptor signal to propagate to the brain, for the transduction
of flight signal in brain, for the flight signal to propagate to muscle, 
and finally for the muscles to actuate the flight. 
The time gap between the flight initiation and the observed flight action is 
the human reaction time. The beam power does not end at the flight initiation; 
it continues until the occurrence of flight action. 

We collected from published literature values of all parameters in the model 
except 1) the volume threshold for flight initiation $v_c$ and 
2) the absorbed beam power density $P_d^{(0)}$. In exposure tests 
$P_d^{(0)}$ is not directly measurable. 
The occurrence time of flight action $t_F$ is reliably measured. 
In our model, $t_F$ is a strictly decreasing function of $P_d^{(0)}$. 
This monotonic trend allows using $t_F$ instead of $P_d^{(0)}$ to parameterize 
the thermal injury risk simulation. 
To get around the unknown volume threshold $v_c$, we normalized the independent 
variables in the formulation. 
In the depth direction, the length scale $z_s$ is naturally given by the penetration 
depth of the electromagnetic frequency. 
The time scale $t_s$ is based on the heat diffusion and the depth scale since the heat 
diffusion is significant only in the depth direction. 
The lateral length scale $r_s$ is derived from the unknown volume threshold 
such that after the scaling, the normalized volume threshold is $\pi$ 
(the volume of a cylinder of height 1 and radius 1). 
Since the volume threshold $v_c$ is unknown, the lateral length scale $r_s$ is 
also unknown. When the physical beam radius is specified as a given 
multiple of the unknown lateral scale $r_s$, 
there is no unknown parameters in the normalized formulation. 
The setting allows simulations in the normalized formulation to 
compute the thermal damage parameter $\Omega $ and use it to classify 
the 1st-degree burn ($\Omega \ge 0.52$), the 2nd-degree burn ($\Omega \ge 1$) 
and the 3rd-degree burn ($\Omega \ge 10^4$). 
In simulations of thermal injury risk, $\Omega $ is parameterized by 
the observed time of flight action $t_F$. The absorbed beam power 
density $P_d^{(0)}$ is hidden in the observed $t_F$. 
This corresponds well to the situation of real exposure tests in which $t_F$ 
is observed but $P_d^{(0)}$ is not measurable. 
Our simulations reveal the conclusions below. 
\begin{itemize}
\item At a fixed physical beam radius, when the observed time of flight action is larger, the underlying absorbed beam power density is smaller, the maximum temperature 
is smaller and the thermal damage parameter is smaller. 
\item At a fixed time of flight action, when a beam of larger radius is used in exposure tests, 
a larger skin area is heated, the activated depth needed for reaching flight initiation is smaller
 and the underlying absorbed beam power density needed is smaller, which produces
a smaller thermal damage parameter.
\item At a fixed observed time of flight action, a shorter human reaction time 
gives more time for electromagnetic heating to reach flight initiation and 
at the same time reduces the duration of undesired additional heating after the flight initiation. 
Both of these two factors contribute to a smaller thermal damage parameter. 
\item At a fixed observed time of flight action, when the skin baseline temperature 
is higher or the nociceptor activation temperature is lower, 
the temperature difference from the baseline to the activation threshold is smaller, 
the underlying absorbed beam power density needed for reaching flight initiation
is smaller and the corresponding thermal damage parameter is smaller. 
\item At a fixed observed time of flight action, when the skin heat conductivity is larger
or the volumetric heat capacity is lower, the heat diffusion is more 
effective in smoothing out a temperature gradient in the depth. As a result, 
the activated depth needed for flight initiation is reached with 
a smaller maximum temperature, 
which yields a smaller thermal damage parameter. 
\item At a fixed observed time of flight action, when the penetration depth of 
the electromagnetic frequency into skin is larger (i.e. the skin absorption
coefficient for the electromagnetic frequency is smaller), 
the heating in the depth is more uniform, the activated depth needed for flight initiation 
is reached with a smaller maximum temperature, which lowers 
the thermal damage parameter. 
\item At a fixed observed time of flight action, when the threshold on the activated 
skin volume for flight initiation is smaller, the underlying absorbed beam power density 
needed for reaching the volume threshold is lower, 
which produces a smaller thermal damage parameter. 
\item
Although a smaller volume threshold and a 
larger penetration depth have qualitatively similar effects, 
they have different mechanism and different effectiveness. 
A smaller volume threshold translates to a smaller heating beam power density 
needed for flight initiation. It reduces the additional heating in the time period 
from the flight initiation to the end of beam power at flight action, which 
is more direct and effective in lowering the thermal damage parameter.
In comparison, a larger penetration depth makes the electromagnetic heating
more uniform. It does not directly reduce the heating. 
It lowers the thermal damage parameter by producing a smaller maximum 
skin temperature at flight initiation and at flight action. 
The mechanism of a larger penetration depth is less effective 
in lowering the thermal damage parameter.
\end{itemize}

\clearpage
\noindent{\bf \large Acknowledgement and disclaimer}

\noindent \indent
The authors acknowledge the Joint Intermediate Force Capabilities Office of U.S. Department of Defense and the Naval Postgraduate School for supporting this work. The views expressed in this document are those of the authors and do not reflect the official policy or position of the Department of Defense or the U.S. Government. 

%%%
%%%% References
%%%

%%%
%%%
%%%

\clearpage
\begin{thebibliography}{hzhou}
%
\bibitem{Romanenko_2017}
Romanenko, S., R. Begley, A. R. Harvey, L. Hool, and V. P. Wallace (2017) The interaction between electromagnetic fields at megahertz, gigahertz and terahertz frequencies with cells, tissues and organisms: risks and potentials. Journal of the Royal Society Interface 14: 20170585 

\bibitem{Zhadobov_2011}
Zhadobov, M., N. Chahat, R. Sauleau, C. Le Quement, and Y. Le Drean (2011) Millimeter-wave interactions with the human body: state of knowledge and recent advances. International Journal of Microwave and Wireless Technologies 3:237-247 

\bibitem{Quement_2014}
Le Quement, C., C. N. Nicolaz, D. Habauzit, M. Zhadobov, R. Sauleau, and Y. Le Drean (2014) Impact of 60-GHz millimeter waves and corresponding heat effect on   endoplasmic reticulum stress sensor gene expression. Bioelectromagnetics 35:444-451 

\bibitem{Nelson_2000}
Nelson, D.A., M. T. Nelson, T. J. Walters and P. A. Mason (2000) Skin heating effects of millimeter-wave irradiation – thermal modeling results. IEEE Transactions on Microwave Theory and Techniques 48:2111-2120 

\bibitem{Walters_2000}
Walters, T. J., D. W. Blick, L. R. Johnson, E. R. Adair, and K. R. Foster (2000) Heating and pain sensation produced in human skin by millimeter waves: comparison to a simple thermal model. Health Phys 78:259-267 

\bibitem{Foster_2010}
Foster, K. R., H. Zhang and J. M. Osepchuk (2010) Thermal response of tissues to millimeter waves: implications for setting exposure guidelines. Health physics 99(6):806-810 

\bibitem{Wang_2020}
Wang, H. , Burgei, W. and Zhou, H. (2020) 
Non-Dimensional Analysis of Thermal Effect on Skin Exposure to 
an Electromagnetic Beam. {\em American Journal of Operations Research}, 
{\bf 10}, 147-162. 
doi: 10.4236/ajor.2020.105011.
%
\bibitem{Cazares_2019}
Cazares, S.M., Snyder, J.A., Belanich, J., Biddle, J.C., Buytendyk, A.M., 
Teng, S.H.M. and O’Connor, K. (2019) 
Active Denial Technology Computational Human Effects End-to-End Hypermodel 
for Effectiveness (ADT CHEETEH-E). 
{\em Human Factors and Mechanical Engineering for Defense and Safety}, 
{\bf 3}, Article No. 13. 
doi:10.1007/s41314-019-0023-7
%
\bibitem{Duck_1990}
Duck, F.A. (1990) 
Physical Properties of Tissue: A Comprehensive Reference Book. 
Academic Press, London.
%
\bibitem{Xu_2010}
Xu, F., Lin, M., Lu, T.J. (2010)
Modeling skin thermal pain sensation: Role of non-Fourier thermal behavior 
in transduction process of nociceptor. 
{\em Computers in Biology and Medicine}, {\bf 40}(5):478-86. 
doi:10.1016/j.compbiomed.2010.03.002
%
\bibitem{Henriques_1947}
Henriques, F.C., Moritz, A.R. (1947) 
Studies of Thermal Injury: I. The Conduction of Heat to and through Skin and 
the Temperatures Attained Therein. A Theoretical and an Experimental Investigation. 
{\em The American Journal of Pathology}, {\bf 23}(4):530-49. 
PMID: 19970945
%
\bibitem{Elkins_1973}
Elkins, W., and Thomson, J. G. (1973) 
Instrumented thermal manikin. 
Acurex Corporation. Aerotherm Division Report No. AD-781, p. 176.
%
\bibitem{Walters_2020}
Walters T.J., Blick D.W., Johnson L.R., Adair E.R., Foster K.R. (2000) 
Heating and Pain Sensation Produced in Human Skin by Millimeter Waves: Comparison to a Simple Thermal Model. {\em Health Physics}, {\bf 78}(3): 259–267.
doi:10.1097/00004032-200003000-00003
%
\bibitem{Olesen_1982}
Olesen, B.W. (1982) 
Thermal comfort. Technical Review, No. 2, Bruel \& Kjaer, Nærum, Denmark
%
\bibitem{Tillman_1995}
Tillman, D.B., Treede, R.D., Meyer, R.A. and Campbell, J,N. (1995)
Response of C fibre nociceptors in the anaesthetized monkey to heat stimuli: 
estimates of receptor depth and threshold. 
{\em The Journal of physiology}, {\bf 485} (Pt 3), 753-65.
doi: 10.1113/jphysiol.1995.sp020766
%
\bibitem{Dykiert_2012}
Dykiert, D., Der, G., Starr, J. M., and Deary, I. J. (2012) 
Sex differences in reaction time mean and intraindividual variability across the life span. 
{\em Developmental Psychology}, {\bf 48}(5), 1262–1276.
doi 10.1037/a0027550
%
\bibitem{Woods_2015}
Woods, D.L., Wyma, J.M., Yund, E.W., Herron, T.J. and Reed, B. (2015)
Factors influencing the latency of simple reaction time. 
{\em Front Hum Neurosci}, {\bf 9}, Article 131. 
doi: 10.3389/fnhum.2015.00131
%
\bibitem{Pearce_2010}
Pearce, J. A. (2010)
Models for Thermal Damage in Tissues: Processes and Applications. 
{\em Critical Reviews in Biomedical Engineering}, {\bf 38}(1): 1–29.
doi: 10.1615/critrevbiomedeng.v38.i1.20
%
\bibitem{Pearce_2017}
Pearce, J. A. , Diller K. R. and Valvano, J. W. (2017)
Bioheat Transfer, in {\em CRC Handbook of Thermal Engineering Second Edition} ed. Raj P. Chhabra (Boca Raton: CRC Press)
%
\bibitem{Orgill_1998}
Orgill, D.P., Solari, M.G., Barlow, M.S., O'Connor, N.E. (1998) 
A finite-element model predicts thermal damage in cutaneous contact burns.
{\em J Burn Care Rehabil}, {\bf 19}(3):203-9. 
doi: 10.1097/00004630-199805000-00003 
%
% ZZZZ
%
%\bibitem{Xu_2008}
%Xu, F., Wen, T., Lu, T. J., and Seffen, K. A. (2008) 
%Modeling of Nociceptor Transduction in Skin Thermal Pain Sensation. 
%{\em ASME. J Biomech Eng}, {\bf 130}(4): 041013. 
%doi:10.1115/1.2939370
%

\bibitem{Topfer_2015}
Topfer, F. and Oberhammer, J. (2015)
Millimeter-wave Tissue Diagnosis: The most Promising Fields for Medical Applications.
{\em IEEE Microwave Magazine}, {\bf 16}(4), 97-113.
%
\bibitem{WBZ_2020}
Wang, H., Burgei, W.A. and Zhou, H. (2020) 
A concise model and analysis for heat-induced withdrawal reflex caused by 
millimeter wave radiation.
{\em American Journal of Operations Research}, {\bf 10}, 31-81. 
%
\bibitem{WBZ_2018uncer}
Wang, H., Burgei, W.A. and Zhou, H. (2018) 
Dose-Injury Relation as a Model for Uncertainty Propagation from Input Dose to Target Dose.
{\em American Journal of Operations Research}, {\bf 8}, 360-385. 
%
\bibitem{Paschotta_2008}
Paschotta, R. (2008) 
Article on ``Effective Mode Area'' in the Encyclopedia of Laser Physics and Technology, 
October 2008, Wiley-VCH, ISBN 978-3-527-40828-3. 
%
%
\end{thebibliography}
\end{document}